\documentclass[lettersize,journal]{IEEEtran}

\usepackage[dvips]{graphicx}
\usepackage[cmex10]{amsmath}
\usepackage{upgreek}
\usepackage{amsthm}
\usepackage{booktabs}
\usepackage{epsfig}
\usepackage{latexsym}
\usepackage{multirow}
\usepackage{stfloats}
\usepackage{epstopdf}
\usepackage{color}  
\usepackage{tabularx} 
\usepackage{amssymb}
\usepackage{enumerate}
\usepackage{enumitem}
\graphicspath{{./Figures/}}
\usepackage{color}
\usepackage{bbm}
\usepackage{bm}
\usepackage{cite}
\usepackage[tight,footnotesize]{subfigure}
\usepackage{balance}
\usepackage{mathrsfs}
\usepackage{verbatim}
\usepackage{dsfont}
\usepackage{verbatim}
\usepackage{tikz}
\usepackage{diagbox}
\usepackage{caption}
\allowdisplaybreaks[4]
\usepackage[framemethod=tikz]{mdframed}
\usepackage{multicol}
\usepackage{environ}
\usepackage{tikz}
\usepackage{algorithm}
\usepackage{algpseudocode}

\usepackage{subfig}
\usepackage{subcaption} 
\begin{document}

\title{Diffusion-Driven Terahertz Air–Ground Communications under Dynamic Atmospheric Turbulence
%AI-Empowered Terahertz Air-Ground Communications in Dynamic Turbulence Medium via Fluid Dynamic-Informed Modeling
}

\author{Jinhao Yi,~\IEEEmembership{Graduate Student Member,~IEEE}, Weijun Gao,~\IEEEmembership{Member,~IEEE}, Chong Han,~\IEEEmembership{Senior~Member,~IEEE}
\thanks{
Jinhao~Yi and Weijun~Gao are with the Terahertz Wireless Communications (TWC) Laboratory, Shanghai Jiao Tong University, Shanghai 200240 China (email:~\{jinhao.yi, gaoweijun\}@sjtu.edu.cn). 

Chong~Han is with the Terahertz Wireless Communications (TWC) Laboratory, Cooperative Medianet Innovation Center (CMIC), School of Information Science and Electronic Engineering, Shanghai Jiao Tong University, Shanghai 200240 China (email:~chong.han@sjtu.edu.cn). 
}
}

%\author{IEEE Publication Technology,~\IEEEmembership{Staff,~IEEE,}
        % <-this % stops a space
%\thanks{This paper was produced by the IEEE Publication Technology Group. They are in Piscataway, NJ.}% <-this % stops a space
%\thanks{Manuscript received April 19, 2021; revised August 16, 2021.}}

% The paper headers
%\markboth{Journal of \LaTeX\ Class Files,~Vol.~14, No.~8, August~2021}%
%{Shell \MakeLowercase{\textit{et al.}}: A Sample Article Using IEEEtran.cls for IEEE Journals}

% Remember, if you use this you should call \IEEEpubidadjcol in the second
% column for its text to clear the IEEEpubid mark.

\maketitle

\begin{abstract}
The ever-increasing demand for ultra-high data rates in space-air-ground integrated networks (SAGINs) has rendered terahertz (THz) communications a promising technology owing to its exceptionally broad and continuous spectrum resources. Nevertheless, in air-ground (AG) scenarios, the high mobility of aircraft induces intense and rapidly fluctuating turbulence, leading to additional propagation loss that is often overlooked in existing studies. To bridge this gap, this paper presents an AI-empowered THz AG communication framework that explicitly models turbulence-induced attenuation through fluid dynamics and integrates it into an adaptive optimization paradigm for communication performance enhancement. Specifically, a fluid-dynamics-informed attenuation model is established to characterize aircraft-generated turbulence and quantify its impact on THz signal propagation. Building upon this model, a joint power-attitude optimization problem is formulated to adaptively allocate transmit power and adjust aircraft attitude for maximizing link capacity. The optimization problem is efficiently solved using a diffusion-based algorithm that learns the nonlinear relationship between flight configuration and turbulence-induced attenuation. Comprehensive numerical evaluations demonstrate that the turbulence-induced attenuation ranges from 18 to 28~$\textrm{dB}$ under attacking angles between $-10^{\circ}$ and $10^{\circ}$ at 0.7 Mach, verifying the pronounced impact of aircraft-induced turbulence on THz propagation. Furthermore, the proposed framework attains an average capacity of 11.241~\textrm{bps/Hz}, substantially outperforming existing strategies by 22.8\% and 66.5\%, and approaching approximately 98\% of the theoretical capacity limit. These findings underscore the critical role of coupling fluid-dynamic modeling with AI-assisted optimization for turbulence modeling, thereby paving the way for reliable, high-capacity THz air-ground links in future SAGIN deployments.
\end{abstract}

\begin{IEEEkeywords}
Terahertz (THz) communications, Space-air-ground integrated network (SAGIN), Atmospheric turbulence, Fluid dynamics, Machine learning.
\end{IEEEkeywords}

\section{Introduction}
\IEEEPARstart{W}{ith} the development of next-generation wireless systems~\cite{saad2019vision,yang20196g,dang2020should}, seamless connectivity, ubiquitous coverage, and ultra-wide-area broadband access are expected to be provided anytime and anywhere~\cite{zhao2019uav,chen2020securing,zhu2020millimeter,zhou2021gateway}. To meet these ambitious requirements, the integration of terrestrial, aerial, and space segments into a space-air-ground integrated network (SAGIN) has been widely recognized as a key paradigm~\cite{xiao2024space,david20186g,cui2022space,zhou2023aerospace}. Within this framework, air-ground (AG) communication plays a pivotal role by connecting airborne platforms with terrestrial nodes. For instance, aircraft such as unmanned aerial vehicles (UAVs) can function as base stations or relays, establishing line-of-sight (LoS) links that mitigate shadowing from obstacles and enhance spectrum efficiency~\cite{zeng2016wireless,geraci2022will}. Their mobility and flexible deployment enable rapid adaptation to dynamic traffic demands, making them particularly valuable in scenarios such as disaster response, temporary high-capacity events, and remote area coverage~\cite{liu2021uav,liu2021novel,huang2024energy}.

To satisfy the ever-increasing demand for ultra-high data rates in these dynamic AG links, Terahertz~(THz) communication has emerged as a promising technology for the next-generation wireless systems~\cite{han2022terahertz,jiang2024terahertz}. Spanning frequencies from $0.1-10~\textrm{THz}$, the THz band offers more contiguous spectrum resources and enables $10-100$ times faster achievable data rates compared with conventional spectrum, thereby addressing the demand for high capacity in airborne links~\cite{chen2021terahertz,petrov2020ieee}. Furthermore, the THz band outperforms free-space optical (FSO) communications with more relaxed alignment requirements and improved robustness to sunlight and fog interference~\cite{gao2025seamless}. These advantages position THz communication as a robust and compelling candidate for reliable, high-speed links in AG integrated networks~\cite{mao2022terahertz, alqaraghuli2023road}.

Despite these aforementioned advantages, there still exist challenges hindering the deployment of THz wireless communication technology in AG communications. One of the fundamental topics lies in accurately THz channel modeling in AG communications~\cite{wang20226g}. Existing studies have made considerable progress in THz AG channel modeling. In~\cite{cui2022cluster,cui2019measurement,an2022measurement}, the impact of terrain-induced multipath effects on AG THz communication channels is investigated. The authors in~\cite{li2021ray,jornet2011channel,han2022molecular} focus on LoS signal attenuation, considering factors such as free-space path loss, molecular absorption due to water vapor, and scattering effects from raindrops and dust. However, these models generally assume a static propagation medium and overlook the dynamic nature of atmospheric turbulence, which may introduce additional attenuation for THz AG communication. Recent studies have investigated the impact of turbulence on THz wave propagation. In~\cite{dordova2010calculation}, the turbulence-induced attenuation is discovered to be proportional to $f^{7/6}$, where $f$ represents the frequency.~\cite{ma2015experimental} examines the effects of turbulence on THz links experimentally under controlled lab conditions.~\cite{gao2024attenuation} and~\cite{cang2019impact} investigate the phase fluctuations and turbulence-induced loss of spatial coherence of THz signals. These studies focus on general communication scenarios and analyze the statistical effect of turbulence on electromagnetic wave propagation. Hence, they often employ empirical models such as the Hufnagle-Valley model~\cite{stotts2022bit}, where the statistical effect is assumed to depend only on height. In contrast, in AG communication scenarios involving high-speed aircraft, the speed and attitude of the aircraft directly influence the surrounding flow field. As a result, existing models generally fail to capture the dynamic coupling between flight configuration and channel characteristics, and have yet to explore how adaptive adjustment of the flight configuration could be leveraged to enhance communication performance.

Motivated by these limitations, we develop an AI-empowered THz AG communication framework that captures these dynamic turbulence effects via fluid dynamics and integrates them into an adaptive optimization scheme for enhanced communication performance. Specifically, we develop a turbulence-aware system model that leverages fluid dynamics to capture the impact of aircraft-induced turbulence on THz signal propagation. Based on this physical fundamental, we formulate a joint power-attitude optimization problem to maximize THz communication capacity by adjusting transmit power allocation and aircraft attitude under turbulence conditions. Furthermore, to efficiently solve the optimization problem with low computational complexity, we introduce a diffusion-based surrogate model that learns the mapping between flight parameters and turbulence effects. This model provides lightweight but accurate predictions of turbulence-induced attenuation, enabling efficient real-time estimation and facilitating faster optimization of THz communication performance. The specific contributions of this paper are summarized as follows.
\begin{itemize}
\item \textbf{We propose a fluid dynamics-based system model that incorporates dynamic turbulent attenuation driven by aircraft configurations variations for THz aircraft communications.} To accurately capture the impact of turbulence on aircraft communications, the system model integrates the Reynolds-Averaged Navier-Stokes (RANS) equations and the SST $k$-$\omega$ turbulence model~\cite{menter2003ten} with the varying flight configurations to characterize the spatial and temporal distributions of temperature, pressure, turbulent kinetic energy and energy dissipation rate in turbulence fields. After resolving these turbulence fields, we directly map these turbulence characteristics to the corresponding turbulence-induced attenuation and derive an accurate closed-form expression that explicitly links flight configurations with turbulent attenuation.

\item \textbf{We formulate a joint power-attitude optimization problem to maximize THz communication capacity in dynamic flight environments.}
Building on the turbulence attenuation model, we formulate an optimization problem that jointly considers transmission power allocation and aircraft attitude adjustment to enhance channel capacity under dynamic flight conditions. To make this nonconvex problem tractable, it is decomposed into two interdependent subproblems: transmission power allocation with fixed attitude~\textbf{Q1}, and attitude adjustment with fixed transmit power~\textbf{Q2}. The closed-form solution of~\textbf{Q1} can be derived analytically, providing an explicit power allocation strategy that serves as an input for subsequent attitude optimization~\textbf{Q2}.

\item \textbf{We develop a diffusion-based surrogate model to efficiently solve the attitude optimization subproblem by efficiently
resolving the THz turbulent attenuation and enable iterative joint optimization.}
Instead of repeatedly solving
the fluid dynamics equations for~\textbf{Q2}, the surrogate model learns the nonlinear mapping from flight configurations to the corresponding turbulence attenuation. Moreover, it is seamlessly integrated into the joint optimization framework, where it functions as a fast turbulence predictor to support attitude optimization in~\textbf{Q2}.
By alternating between the two subproblems, the framework adaptively refines both power allocation and attitude adjustment, progressively improving communication performance under dynamic flight conditions.

\item \textbf{We evaluate the impact of dynamic flight configurations on THz aircraft communications via the proposed fluid dynamics-based turbulence model and optimization framework.} The effects of aircraft speed and attitude on THz turbulent attenuation and achievable capacity are systematically examined via numerical simulations. The results show that turbulent attenuation is non-negligible, ranging from $18-28~\textrm{dB}$ under typical flight conditions. Moreover, the attenuation increases with higher speed and varies substantially with different attitudes. We further assess the proposed joint optimization framework, where the optimized strategy achieves an average capacity of $11.241~\textrm{bps}/\textrm{Hz}$, approaching the expert benchmark with only a marginal gap and outperforming random and fixed strategies that remain below $10~\textrm{bps}/\textrm{Hz}$. 
\end{itemize}

The remainder of the paper is structured as follows. In Sec.~\ref{System_Model}, we introduce the system model and develop the fluid dynamics-informed model at THz band that captures the additional turbulence-induced propagation attenuation. In Sec.~\ref{formulation}, we formulate the optimization problem and propose the iterative solution by developing the diffusion-based surrogate model that enables efficient optimization for THz communication capacity. Numerical results are demonstrated in Sec.~\ref{numerical} and the paper is concluded in Sec.~\ref{conclusion}.

\begin{figure*}[t]
\centerline{\includegraphics[width=0.8\textwidth]{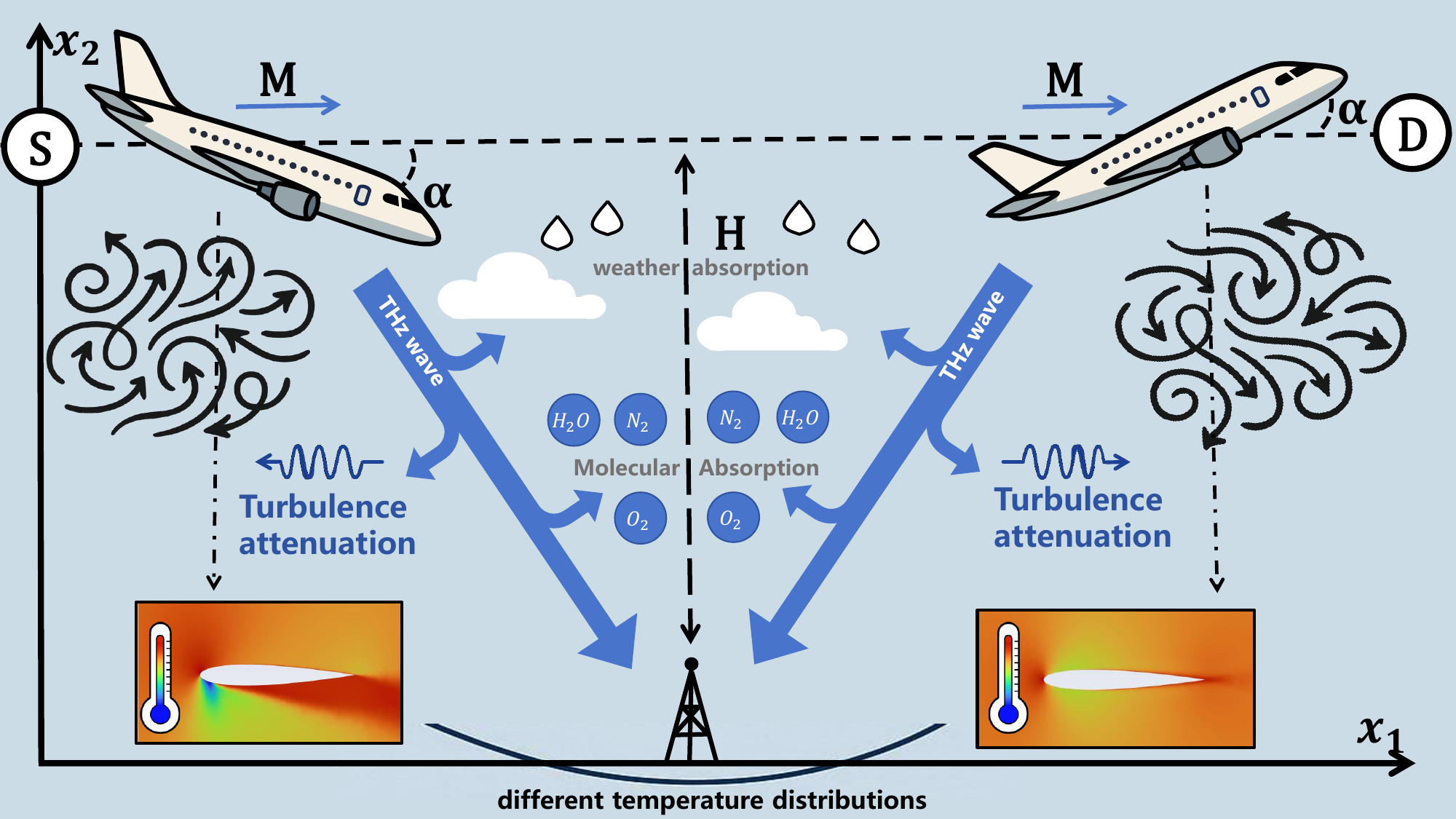}}
\captionsetup{font={footnotesize}}
\caption{System model.}
\label{fig_environment}
\end{figure*}

\section{System Model}
\label{System_Model}
We consider an AG communication scenario involving a high-speed aircraft and a base station as illustrated in Fig.~\ref{fig_environment}. The aircraft flies along a predetermined trajectory from starting node $S$ to destination node $D$ in a varying speed denoted as $M(t)$ in Mach number. Meanwhile, THz communication is continuously established between the aircraft and the stationary base station on the ground. simplify the analysis, we assume the aircraft  flying over a horizontal flight path $L$ at a fixed altitude $H$, and the ground-based station, denoted as $BS$, is located at the midpoint of the flight path. Moreover, we assume the THz AG communication channel from aircraft to $BS$ is dominated by the LoS path. We consider a two-dimensional Cartesian coordinate system with the horizontal axis along the $x_1$-direction and the vertical axis along the $x_2$-direction. In this system, the positions with $S$, $D$, and $BS$ are located at $(0, H)$, $(L, H)$, and $\left(\frac{L}{2}, 0\right)$, respectively, and the time-varying position of the aircraft can be expressed as $\left(x\left(t\right), H\right),\ 0\leq t \leq T $, where $T$ represents the maximum allowable time for the aircraft to travel from $S$ to $D$. In this scenario, turbulence induced by the high-speed aircraft causes non-negligible attenuation in THz communications, which must be accounted for in the channel model. 
According to~\cite{tatarski2016wave,wu2021reliable,beland1993propagation}, the turbulence-induced attenuation is determined by the turbulence fields, i.e., the spatial distributions of temperature $T$, pressure $P$, turbulent kinetic energy $\mathcal{E}$ and energy dissipation rate $\omega$. 
However, the turbulence fields surrounding the aircraft, such as the spatial distributions of $T$, are influenced by flight speed $M$ and attitude $\alpha$. To capture this, we propose a fluid dynamics-based approach, where the turbulence fields around the aircraft are first computed in Sec.~\ref{subA}, and the resulting turbulence-induced attenuation, which is denoted as~$L_\mathrm{turb}^{M(t),\alpha(t)}$, is characterized in Sec.~\ref{subB} via the computed turbulence fields. These computations are then integrated into the overall channel model.

\subsection{Fluid Dynamic-informed Modeling of Turbulence 
Fields}
\label{subA}
Since $L_\mathrm{turb}^{M(t),\alpha(t)}$ is determined by the turbulence fields, we develop a fluid dynamic-informed high-fidelity model to characterize the turbulence fields around the aircraft, incorporating its speed $M(t)$ and attitude $\alpha(t)$.
Fundamentally, the physical fields of atmospheric turbulence around the aircraft are governed by Navier-Stokes (N-S) equations and energy equations, which are expressed as
\begin{equation} 
-\frac{\partial P}{\partial x_i} = 
\frac{\partial (\rho u_i)}{\partial t} + \frac{\partial (\rho u_j u_i)}{\partial x_j} - \frac{\partial}{\partial x_j} \left[ \mu \left( \frac{\partial u_i}{\partial x_j} + \frac{\partial u_j}{\partial x_i} \right) \right],
\label{eq_ns} 
\end{equation}

\begin{equation}
\begin{aligned}
&\rho c_p \left( \frac{\partial T}{\partial t} + u_i \frac{\partial T}{\partial x_i} \right) = \frac{\partial}{\partial x_j} \left( \lambda \frac{\partial T}{\partial x_j} \right) + \Phi,
\end{aligned}
\label{eq_energy}
\end{equation}
where $i, j \in \{1, 2\}$ denotes two arbitrary dimensions in the coordinate. Parameters $\rho, \mu, c_p, \lambda$ represent the constants of fluid density, molecular viscosity, heat capacity, and thermal conductivity, respectively, and $u_i$ denote the velocity components in the \( x_i \)-direction. $\Phi$ denotes the viscous dissipation related to the different kinds of fluid.
Analytical and numerical solutions directed to N-S equations are challenging and time-consuming, especially for turbulence fields.
We therefore apply the Reynolds-Averaged Navier-Stokes (RANS) approach to efficiently solve the statistics of the turbulence field based on N-S equations and energy equations, where only the time-averaged velocity components are considered. By applying the averaging process, the RANS equations and energy equations are given by 
\begin{equation}
\begin{aligned}
-\frac{\partial P}{\partial x_i} & = 
\frac{\partial (\rho \overline{u_i})}{\partial t} + \frac{\partial}{\partial x_j} (\rho \overline{u_i} \overline{u_j}) \\ 
& - \frac{\partial}{\partial x_j} \left[ \mu \left( \frac{\partial \overline{u_i}}{\partial x_j} + \frac{\partial \overline{u_j}}{\partial x_i} \right) - \Omega \right] ,
\end{aligned}
\label{eq6}
\end{equation}

\begin{equation}
\begin{aligned}
&\rho c_p \left( \frac{\partial T}{\partial t} + u_i \frac{\partial T}{\partial x_i} \right) = \frac{\partial}{\partial x_j} \left( \lambda \frac{\partial T}{\partial x_j} \right) + \Phi',
\end{aligned}
\label{eq_7}
\end{equation}
where $\overline{u_i}$ and $\overline{u_j}$ are the time-averaged velocity components. $\Omega$ and $\Phi'$ are the additional correction terms of turbulence called Reynolds stress tensor and turbulent viscous dissipation, respectively, and are further modeled by the turbulent viscosity $\mu_t$ and are given by

\begin{equation}
\Omega = \mu_t \left( \frac{\partial u_i}{\partial x_j} + \frac{\partial u_j}{\partial x_i} \right) - \frac{2}{3} \left( \rho \mathcal{E} + \mu_t \frac{\partial u_i}{\partial x_i} \right) \delta_{ij},
\label{ns_eq}
\end{equation}

\begin{equation}
\Phi'= \mu_t \left( \frac{\partial u_i}{\partial x_j} + \frac{\partial u_j}{\partial x_i} \right) \frac{\partial u_i}{\partial x_j} + \mu_t \left( \frac{\partial u_j}{\partial x_j} + \frac{\partial u_i}{\partial x_j} \right) \frac{\partial u_i}{\partial x_j}
\end{equation}
where $\mathcal{E}$ is the turbulent kinetic energy. It is noted that $\mu_t$ is the effective turbulent viscosity that differs from the constant molecular viscosity \( \mu \).
To model $\mu_t$, the Shear Stress Transport (SST) model~\cite{menter2003ten} is widely used in aerospace applications, where $\mu_t$ is modeled as 
\begin{equation}
\mu_t = \rho \cdot \frac{\mathcal{E}}{\max(\omega, F_2 S)},
\end{equation}
where $F_2$ is a predetermined blending function, $S$ is the strain rate magnitude, and $\omega$ is the energy dissipation rate. $\mathcal{E}$ and $\omega$ are further governed by
\begin{equation}
\begin{aligned}
\frac{\partial (\rho \mathcal{E})}{\partial t} + \frac{\partial (\rho u_j \mathcal{E})}{\partial x_j} &= P_{\mathcal{E}} - \beta_1 \rho \mathcal{E} \omega \\&+ \frac{\partial}{\partial x_j} \left[ \left( \mu + \sigma_{\mathcal{E}} \mu_t \right) \frac{\partial \mathcal{E}}{\partial x_j} \right],
\label{eq_k}
\end{aligned}
\end{equation}

\begin{equation}
\begin{aligned}
\frac{\partial (\rho \omega)}{\partial t} &+ \frac{\partial (\rho u_j \omega)}{\partial x_j} =  \frac{\partial}{\partial x_j} \left[ \left( \mu + \sigma_\omega \mu_t \right) \frac{\partial \omega}{\partial x_j} \right] \\
&\beta_2 \frac{\omega}{\mathcal{E}} P_{\mathcal{E}} - \beta_3 \rho \omega^2  + 2(1 - F_1) \rho \sigma_{\omega} \frac{1}{\omega} \frac{\partial \mathcal{E}}{\partial x_j} \frac{\partial \omega}{\partial x_j}, 
\end{aligned}
\label{eq_k2}
\end{equation}
where $P_{\mathcal{E}}$ is the production term of turbulent kinetic energy, $\beta_1$,  $\beta_2$,  $\beta_3$, $\sigma_{\mathcal{E}}$, and $\sigma_\omega$ are constants, and $F_1$ is another predetermined blending function. When $F_1 = 1$, the SST model is more suitable for near-wall regions due to its sensitivity to boundary layer effects. When $F_1 = 0$, the model performs better in the free-stream region. 
Furthermore, the temperature 
Following the RANS approach and SST model in~\eqref{eq6}-\eqref{eq_k2}, the boundary conditions are subsequently defined to enable the computation of the turbulence fields. In our model, the boundary conditions are defined by the flight speed $M$ and attitude $\alpha$. Different combinations of $\{M(t),\alpha(t)\}$ at different times $t$ lead to distinct computation results of turbulence fields and further result in distinct $L_\mathrm{turb}^{M(t),\alpha_(t)}$.

Given the governing equations and boundary conditions, the turbulence fields including the spatial distributions of temperature, pressure, turbulent kinetic energy and energy dissipation rate under the combination of $\{M(t),\alpha(t)\}$ at time $t$, denoted as $T^{M(t),\alpha(t)}$, $P^{M(t),\alpha(t)}$, $\mathcal{E}^{M(t),\alpha(t)}$ and $\omega^{M(t),\alpha(t)}$, can all be accurately computed numerically, since the analytical solutions are still not feasible due to the complexity of solving partial differential equations.

Building upon the computed spatial distributions of turbulence fields, in the following, we focus on modeling the impact of turbulence on THz communication in AG channel. Specifically, we derive a closed-form expression for turbulence-induced attenuation, which incorporates the relevant physical quantities of the computed turbulence fields.

\subsection{Modeling of Turbulence-Induced Attenuation Based on Computed Turbulence Fields}
\label{subB}
Based on the computed turbulence fields, the turbulence-induced attenuation $L_\mathrm{turb}^{M(t),\alpha(t)}$ can be characterized in a closed form, which is defined as
\begin{equation}
L_\mathrm{turb}^{M(t),\alpha(t)} = 10 \log \left| 1 - \sqrt{\frac{1}{\alpha(t)} + \frac{1}{\beta(t)} + \frac{1}{\alpha(t) \times \beta(t)}} \right|,
\label{eq12}
\end{equation}
where $\alpha(t)$ and $\beta(t)$ refer to large-scale and small-scale fading parameters at time $t$, and are defined as
\begin{equation}
\alpha(t) = \left[ \exp \left( \frac{0.49 {\sigma(t)}^2}{(1 + 0.18 {D(t)}^2 + 0.56 {\sigma(t)}^{\frac{12}{5}})^{\frac{7}{6}}} \right) - 1 \right]^{-1},
\label{eq14}
\end{equation}

\begin{equation}
\beta(t) = \left[ \exp \left( \frac{0.51 {\sigma(t)}^2 (1 + 0.69 {D(t)}^2 {\sigma(t)}^{\frac{12}{5}})^{\frac{-5}{6}}}{(1 + 0.9 {D(t)}^2 + 0.62 {\sigma(t)}^{\frac{12}{5}})^{\frac{7}{6}}} \right) - 1 \right]^{-1},
\label{eq15}
\end{equation}
where \( D(t) = \sqrt{\frac{\pi fl^2}{2r(t)}} \), and \( l = \frac{\lambda}{\pi} \). The only unknown parameter left is ${\sigma_k^i}^2$, which is a crucial parameter related to turbulence fields. According to previous work~\cite{al2001mathematical}, ${\sigma(t)}^2$ is given by
\begin{equation}
\begin{aligned}
{\sigma(t)}^2 &= 2.25 \left(\frac{2\pi f}{c}\right)^{7/6} (H - h_0)^{5/6} \\&\times\int_{h_0}^H B(t) \left(\frac{h - h_0}{H - h_0}\right)^{5/6} \mathrm dh,
\label{eq_height}
\end{aligned}
\end{equation}
where $B(t,h)$ is determined by turbulence fields but is assumed to be only height-dependent and is modeled empirically in~\cite{al2001mathematical} due to the difficulty of obtaining accurate turbulence fields. In contrast, our work directly computes the turbulence fields based on RANS approach in Sec.~\ref{subA}, enabling a more accurate and physics-based modeling of~$B(t,\vec{r})$, ${\sigma(t)}^2$. According to Tatarskii’s study~\cite{tatarski2016wave}, $B(t,\vec{r})$ is determined by 

\begin{equation}
\begin{aligned}
B(t,\vec{r_j}) &=  \frac{c_0 \left(\sqrt{\mathcal{E}^{M(t),\alpha(t)}_j \cdot \omega^{M(t),\alpha(t)}_j}\right)^{4/3}}{{T_j^{M(t),\alpha(t)}}^2} \\ & \times\left( \frac{\partial \left(T^{M(t),\alpha(t)}_j \left(\frac{1000}{P^{M(t),\alpha(t)}_j}\right)^{0.286}\right)}{\partial h}\right)^2
\end{aligned}
\end{equation}
where $c_0$ is a constant~\cite{beland1993propagation}, $T^{M(t),\alpha(t)}_j$ and $P^{M(t),\alpha(t)}_j$, $\mathcal{E}^{M(t),\alpha(t)}_j$ and $\omega^{M(t),\alpha(t)}_j$ are the temperature, pressure, turbulence kinetic energy and dissipation rate under flight speed $M(t)$ and attitude $\alpha(t)$ at time $t$ and point $\vec{r_j}=\left(x_1^j,x_2^j\right)$, respectively. It is important to note that $T^{M(t),\alpha(t)}_j$, $P^{M(t),\alpha(t)}_j$, $\mathcal{E}^{M(t),\alpha(t)}_j$, and $\omega^{M(t),\alpha(t)}_j$ have all been derived through~\eqref{eq6}-\eqref{eq_k2} and boundary conditions in Sec.~\ref{subA}.
Moreover, for aerospace communications with strong turbulence, the turbulence fields are position-dependent, which means ${\sigma(t)}^2$ is no longer solely height-dependent but also varies with position. Therefore, in our scenario, we generalize~\eqref{eq_height} via path integration, which is given by 
\begin{equation}
\begin{aligned}
{\sigma(t)}^2 &= 2.25 \left(\frac{2\pi f}{c}\right)^{7/6} (H - h_0)^{5/6} \\ &\times \int_{r(t)} B(t,\vec{r})\left(\frac{h - h_0}{H - h_0}\right)^{5/6} \mathrm dr,
\label{eq18}
\end{aligned}
\end{equation}
where $r(t)$ denotes the propagation path at time~$t$. This refined formulation seamlessly incorporates both vertical stratification and horizontal distributions of turbulence, thereby enabling a more accurate and physically consistent evaluation of THz link attenuation under realistic and dynamic flight conditions. Hence, the turbulence-induced attenuation $L_\mathrm{turb}^{M(t),\alpha(t)}$ can be derived and is given at the bottom of this page. It can then be incorporated into the THz AG channel model. 

For ease of exposition, the time horizon \( T \) in our system model is discretized into \( K \) equally-spaced time slots, i.e., $T = k \delta_t$, where \( \delta_t \) denotes the sufficiently small slot length. Thus, the flight trajectory over \( T \) can be approximated by the \( K \)-length sequences \( \{ x_k, H \}_{k=1}^K \), where $x_k$ denotes the flight $x$ coordinate at slot \( k \).  
According to~\cite{jornet2011channel, series2019attenuation}, the THz AG channel is frequency-selective with additive white Gaussian noise (AWGN) whose variance is denoted as $N$. Thus, to obtain the wideband capacity, the entire bandwidth is divided into several sub-bands where $f_i$ denotes the center frequency of the $i$-th sub-band. Since the bandwidth of each sub-band, denoted as $\Delta f$, is sufficiently narrow to approximate the channel as non-selective within each sub-band, the capacity in THz AG channel at time slot~$k$ can be given by
\begin{equation}
C_k=\sum_{i}\Delta f\cdot
\log_2 \left(1 + \frac{P_{Rx}^{k,i}}{N}\right),
\label{capa}
\end{equation}
where $P_{Rx}^{k,i}$ is the received signal power of the $i$-th sub-band at time slot $k$ and is given by~\cite{jornet2011channel, series2019attenuation}
\begin{equation}
   P_{Rx}^{k,i}= \frac{c^2G_{Tx}G_{Rx}P_{Tx}^{k,i}\cdot e^{ -\int_{0}^H \mu_{\mathrm{abs}}(f_i,r_k)dr}}{\left(4\pi r_kf_i\right)^2L_\mathrm{Rain}^{k,i}L_\mathrm{Cloud}^{k,i}L_\mathrm{turb}^{M_k,\alpha_k,i}},
\label{capacity}
\end{equation}
where \( G_{Tx} \) and \( G_{Rx} \) are the Tx and Rx antenna gains, and \( c \) is the speed of light. $r_k$, \( P_{Tx}^k \), \( \mu_\mathrm{abs}(f, r_k) \) are the propagation distance, transmit power and the molecular absorption coefficient at time slot~$k$, respectively, and $L_\mathrm{Rain}^{k,i}$ and $L_\mathrm{Cloud}^{k,i}$ are the path loss due to rain and cloud given in~\cite{kokkoniemi2021channel}.

In addition to $L_\mathrm{Rain}^{k,i}$ and $L_\mathrm{Cloud}^{k,i}$, $L_\mathrm{turb}^{M_k,\alpha_k,i}$ represents the turbulence-induced attenuation for the $i$-th sub-band at the discrete time slot $k$. It is obtained by sampling the continuous-time turbulence loss $L_\mathrm{turb}^{M(t),\alpha(t)}$ in~\eqref{eq:L} at $t = k\delta_t$ and $f=f_i$.
Thus, the capacity $C_k$ in THz AG channel at time slot \( k \) is finally derived.

\begin{figure*}[!b]
\hrulefill
\begin{equation}
\begin{aligned}
L_\mathrm{turb}^{M(t),\alpha(t)} &= 10\log_{10}\Bigg|1-\textrm{sqrt}\Bigg\{
\exp\left(\frac{0.49{\sigma(t)}^2}{(1+0.18{D(t)}^2+0.56{\sigma(t)}^{\frac{12}{5}})^{\frac{7}{6}}}\right)
+\exp\left(\frac{0.51{\sigma(t)}^2(1+0.69{D(t)}^2{\sigma(t)}^{\frac{12}{5}})^{\frac{-5}{6}}}{(1+0.9{D(t)}^2+0.6{\sigma(t)}^{\frac{12}{5}})^{\frac{7}{6}}}\right)\\&
+\left[\exp\left(\frac{0.49{\sigma(t)}^2}{(1+0.18{D(t)}^2+0.56{\sigma(t)}^{\frac{12}{5}})^{\frac{7}{6}}}\right)-1\right]\left[\exp\left(\frac{0.51{\sigma(t)}^2(1+0.69{D(t)}^2{\sigma(t)}^{\frac{12}{5}})^{\frac{-5}{6}}}{(1+0.9{D(t)}^2+0.62{\sigma(t)}^{\frac{12}{5}})^{\frac{7}{6}}}\right)-1\right]
\Bigg\}\Bigg|-2,
\end{aligned}
\label{eq:L}
\end{equation}
where 
\begin{equation}
\begin{aligned}
{\sigma(t)}^2 = \frac{2.25 \left(\frac{2\pi f}{c}\right)^{7/6} }{(H - h_0)^{\frac{-7}{6}} }\int_{r(t)}\frac{c_0 \left(\sqrt{\mathcal{E}^{M(t),\alpha(t)}\cdot \omega^{M(t),\alpha(t)}}\right)^{4/3}}{{T^{M(t),\alpha(t)}}^2}\left( \frac{\partial \left(T^{M(t),\alpha(t)} \left(\frac{1000}{P^{M(t),\alpha(t)}}\right)^{0.286}\right)}{\partial h}\right)^2\left(\frac{h - h_0}{H - h_0}\right)^{\frac{5}{6}} dr.\nonumber
\end{aligned}
\end{equation}
\end{figure*}

\section{Problem Formulation and Iterative Solution} 
\label{formulation}
Given the dynamic environmental attenuation in THz AG channel, our objective is to maximize the total capacity of aircraft from $S$ to $D$. According to~\eqref{capa} and~\eqref{capacity}, the problem can be formulated as
\begin{subequations}
\begin{flalign}
\textbf{Q0}: \underset{\{P^{k,i}_{Tx},M_k,\alpha_k\}}{\text{max}}\quad & 
\sum_{k=1}^{K}C_k=\sum_{k=1}^{K} \sum_{i}\Delta f\cdot \nonumber \\
& \log_2 \left(1 + \frac{A_k^iP^{k,i}_{Tx}}{L_\mathrm{turb}^{M_k,\alpha_k}\cdot N}\right) \label{problem}
\\
\text{s.t.}\quad \quad \quad
& \frac{1}{K} \sum_{k=1}^{K}\sum_{i}P^{k,i}_{Tx}\leq \bar{P},   \label{st1}\\
& P^{k,i}_{Tx} \geq 0, \quad \forall i,\forall k \in \{1, \ldots, K\}, \label{st2}\\
% & M_0 = \sqrt{\left( x_1 - x_0\right)^2} \geq \frac{L}{T}\cdot\delta_t, \label{st3}\\
% & M_k = \sqrt{\left( x_k - x_{k-1}\right)^2} \geq \frac{L}{T}\cdot\delta_t,\nonumber\\
% &\quad \quad \quad \quad \quad \quad
% \forall k \in \{2, \ldots, K\}, 
% \label{st4}\\
&\sum_{k=1}^{K}\sum_{i}M_k\geq \bar{M} = \frac{L}{T}\cdot\delta_t,\label{st7}\\
& M_k \in \mathcal{A}_M, \quad \forall k \in \{1, \ldots, K\}, \label{st5}\\
& \alpha_k \in \mathcal{A}_\alpha, \quad \forall k \in \{1, \ldots, K\},\label{st6}
\end{flalign}
\end{subequations}
where $M_k$, $\alpha_k$ and $L_\mathrm{turb}^{M_k,\alpha_k}$ represent the flight speed in Mach number, the flight attitude and the corresponding turbulent attenuation at time slot~$k$. $P^{k,i}_{Tx}$ is the transmit power of the $i$-th sub-band at time slot $k$. $A_k^i$ represents the other coefficients of received signal power in~\eqref{capacity} and is given as 
\begin{equation}
A_k^i= \frac{c^2G_{Tx}G_{Rx} e^{ -\int_{0}^H \mu_{\mathrm{abs}}(f_i,r_k)dr}}{\left(4\pi r_kf_i\right)^2L_\mathrm{Rain}^kL_\mathrm{Cloud}^k},
\end{equation}
Constraint~\eqref{st1} represents the average transmit power constraints over the limited time period $T$ and limited average power $\bar{P}$. Constraints~\eqref{st7} denotes the speed constraints of aircraft at $S$ and over the flight path $L$, which ensure that the aircraft maintains a realistic and efficient trajectory within the limited time period $T$. Furthermore, constraints~\eqref{st5} and~\eqref{st6} impose feasible domains $\mathcal{A}_M, \mathcal{A}_\alpha$ on both the aircraft’s speed and flight attitude to ensure the safety and stability of the aircraft, as well as the safety of the pilot. For simplicity, $\mathcal{A}_M$ and $\mathcal{A}_\alpha$ in our work are restricted to finite and discrete values. 
Therefore, the optimization problem in~\eqref{problem} is turned into a Mixed-Integer Nonlinear Programming (MINLP) problem, where the decision variables include both the continuous variable of the power allocation~$P^{i,k}_{Tx}$ and the discrete variables of~$M_k$ and~$\alpha_k$. MINLP problems are generally NP-hard, leading that they are computationally expensive and difficult to solve. In the following, we propose an algorithm based on problem decomposition and iteration to overcome the computational difficulties.

\subsection{Decomposition of the Optimization Problem}
Since the objective function in \eqref{problem} exhibits a multiplicative form in terms of $P^{k,i}_{Tx}$ and $L_\mathrm{turb}^{M_k,\alpha_k}$, and the constraints~\eqref{st1}-\eqref{st6}
can be separated into terms dependent on distinct subsets of continuous and discrete variables, the original MINLP problem defined in \eqref{problem} can be decomposed into two subproblems:
\begin{enumerate}[leftmargin=*,label=(\roman*)]
\item \textbf{Power Optimization with fixed flight configuration}:
\begin{subequations}
\begin{flalign}
\textbf{Q1}: \underset{\{P^{k,i}_{Tx}\}}{\text{max}}\quad & 
\sum_{k=1}^{K}C_k=\sum_{k=1}^{K} \sum_{i}\Delta f\cdot \nonumber \\
& \log_2 \left(1 + \frac{A_k^iP^{k,i}_{Tx}}{L_\mathrm{turb}^{M_k^*,\alpha_k^*}\cdot N}\right) \label{sub1} \\
\text{s.t.}\quad  
& \frac{1}{K} \sum_{k=1}^{K}\sum_{i}P^{k,i}_{Tx}\leq \bar{P},   \nonumber\\
& P^{k,i}_{Tx} \geq 0, \quad \forall i,\forall k \in \{1, \ldots, K\}, \nonumber
\end{flalign}
\end{subequations}
where $M_k^*$ and $\alpha_k^*$ denote the fixed flight speed and altitude.

\item \textbf{Flight configuration Optimization with fixed power}:
\begin{subequations}
\begin{flalign}
\textbf{Q2}: \underset{\{M_k,\alpha_k\}}{\text{max}}\quad & 
\sum_{k=1}^{K}C_k=\sum_{k=1}^{K} \sum_{i}\Delta f\cdot \nonumber \\
& \log_2 \left(1 + \frac{A_k^i\hat{P}^{k,i}_{Tx}}{L_\mathrm{turb}^{M_k,\alpha_k}\cdot N}\right) \label{sub2}
\\
\text{s.t.}\quad 
& M_0^2 = \sqrt{\left( x_1 - x_0\right)^2} \geq \frac{L}{T}\cdot\delta_t, \nonumber\\
&\sum_{k=1}^{K}\sum_{i}M_k\geq \bar{M} = \frac{L}{T}\cdot\delta_t,\label{st7}, \nonumber\\
& \alpha_k \in \mathcal{A}_\alpha, \quad \forall k \in \{1, \ldots, K\},\nonumber
\end{flalign}
\end{subequations}
where $\hat{P}^{k,i}_{Tx}$ is the fixed transmit power.
\end{enumerate}

After decomposing~\textbf{Q0}, the original MINLP problem is divided into two subproblems~\textbf{Q1} and~\textbf{Q2}. Specifically, \textbf{Q1} focuses on optimizing the transmit power with a fixed flight configuration, while \textbf{Q2} optimizes the flight configuration with fixed transmit power. In the following, we first present the methods for solving both subproblems.

\subsection{Power Optimization with Fixed Flight Configuration} After the decomposition of the optimization problem, the inner continuous subproblem~\textbf{Q1} with constraints~\eqref{st1} and~\eqref{st2} can be solved as follows.

\it 
Lemma 1 (Constraint Tightness):
In the optimal solution of the continuous subproblem~\textbf{Q1}, the total power constraint~\eqref{st1} should be tight:
\begin{equation}
\frac{1}{K} \sum_{k=1}^K \sum_i P_{Tx}^{i,k} = \bar{P}.
\label{tight}
\end{equation}

Proof: 
\rm We prove by contradiction. Suppose there exists an optimal solution \( \{\tilde{P}_{Tx}^{i,k}\} \) such that:

\begin{equation}
\frac{1}{K} \sum_{k,i} \tilde{P}_{Tx}^{i,k} = \bar{P} - \epsilon, \quad \epsilon > 0.
\end{equation}
Then we can construct a new solution defined as 
\begin{equation}
\acute{P}_{Tx}^{i,k} = \tilde{P}_{Tx}^{i,k} + \frac{\epsilon}{K},
\end{equation}
which ensures the total power constraint~\eqref{st1}: 
\begin{equation}
 \frac{1}{K} \sum_{k,i} \acute{P}_{Tx}^{i,k} = \bar{P}.  
\end{equation}
Since the capacity \( C_k\) in~\eqref{problem} is monotonically increasing with \( {P}_{Tx}^{i,k} \), the new solution $\acute{P}_{Tx}^{i,k}$ ensures a higher objective:
\begin{equation}
    \sum_{k,i} C(\acute{P}_{Tx}^{i,k}) > \sum_{k,i} C(\tilde{P}_{Tx}^{i,k}).
\end{equation}
Therefore, the total power constraint~\eqref{st1} should be tight at optimality. 
$\hfill\blacksquare$

Lemma 1 establishes that the total power constraint should be tight at optimality, meaning that any optimal solution cannot leave unused power. It also simplifies the solution process, as it eliminates the need to consider power deficits or excesses during optimization. Moreover, it ensures the subproblem~\textbf{Q1} supports the water-filling principle.

\it 
Theorem 1 (Generalized Water-Filling with Channel Loss):
For the subproblem defined in~\textbf{Q1} with constraints~\eqref{st1} and~\eqref{st2}, the optimal power allocation $\{\breve{P}_{Tx}^{i,k}\}$ is given as

\begin{equation}
\breve{P}_{Tx}^{i,k} = \left[ \frac{K \Delta f_i}{\lambda \ln 2} - \frac{L_\mathrm{turb}^{M_k,\alpha_k}\cdot N}{A_k^i} \right]^+, \quad \forall k,i,
\label{opt1}
\end{equation}
where \( [\cdot]^+ \triangleq \max(\cdot, 0) \) ensures non-negative power allocation and \( \lambda \) is determined by the total power constraint:
\begin{equation}
\sum_{k,i} \left[ \frac{K \Delta f}{\lambda \ln 2} - \frac{L_\mathrm{turb}^{M_k,\alpha_k}\cdot N}{A_k^i} \right]^+ = K\bar{P}.  
\end{equation}

Proof: 
\rm According to Karush-Kuhn-Tucker conditions, the optimal solution \( \{\breve{P}_{Tx}^{i,k}\} \) to the subproblem~\textbf{Q1} with constraint~\eqref{st1} and the tight constraint~\eqref{tight} should satisfy:
\begin{subequations}
\begin{flalign}
&\dfrac{\Delta f_i/\ln 2}{L_\mathrm{turb}^{M_k,\alpha_k}\cdot N/A_k^i + \breve{P}_{Tx}^{i,k}} = \dfrac{\lambda}{K} - \mu_{k,i},
\label{kkt1}\\
 &\lambda \left( \dfrac{1}{K}\sum\limits_{k,i} \breve{P}_{Tx}^{i,k} - \bar{P} \right) = 0,\label{kkt2}\\
 &\mu_{k,i} \breve{P}_{Tx}^{i,k} = 0, \label{kkt3} \\
 &\dfrac{1}{K}\sum\limits_{k,i} \breve{P}_{Tx}^{i,k} \leq \bar{P}, \quad \breve{P}_{Tx}^{i,k} \geq 0,  \label{kkt4}\\
 &\lambda \geq 0, \quad \mu_{k,i} \geq 0, \label{kkt5} 
\end{flalign} 
\end{subequations}
where $\lambda$ and $\mu_{k,i}$ are the Lagrange multipliers for the total power constraint and the non-negativity constraints, respectively. Thus, from~\eqref{kkt1} and~\eqref{kkt4}, the optimal power allocation \( \{\breve{P}_{Tx}^{i,k}\} \) is solved as 
\begin{equation}
\breve{P}_{Tx}^{i,k} = \left[ \frac{K \Delta f}{\lambda \ln 2} - \frac{L_\mathrm{turb}^{M_k,\alpha_k}\cdot N}{A_k^i} \right]^+, \quad \forall k,i, \nonumber
\end{equation}
where \( \lambda \) is determined by the total power constraint~\eqref{kkt2} and~\eqref{kkt5}. Replacing~\eqref{opt1} into~\eqref{kkt2}, the value of \( \lambda \) is determined as
\begin{equation}
\sum_{k,i} \left[ \frac{K \Delta f}{\lambda \ln 2} - \frac{L_\mathrm{turb}^{M_k,\alpha_k}\cdot N}{A_k^i} \right]^+ = K\bar{P}, \nonumber 
\end{equation}
which completes the proof.$\hfill\blacksquare$

Following the successful resolution of~\textbf{Q1}, the next step is to address~\textbf{Q2}, which necessitates the real-time computation of turbulent attenuation. While the RANS-based computational approach, as discussed earlier, offers high accuracy in characterizing this phenomenon, it is still computationally expensive and time-consuming in the aerospace scenario. To overcome this limitation, we introduce a diffusion-based surrogate model in the following section, designed to efficiently approximate the turbulent attenuation and accelerate the solution of the second subproblem.

\subsection{Flight Configuration Optimization with Fixed Power via Diffusion-Based Surrogate Model}
\label{prediction}
In order to efficiently solve the second subproblem~\textbf{Q2}, we first develop an AI-based surrogate model to address the challenge of accurately and quickly computing the turbulence field.
Prior AI-based algorithms for rapid turbulent attenuation estimation~\cite{wang2016using, saha2022turbulence} often assume that the distributions of temperature and pressure are known in advance. However, for THz aircraft communications, the real-time distributions of temperature and pressure are dynamically affected by the speed and attitude of aircraft, leading to turbulence fields that vary both spatially and temporally. We model it as a generative problem, making diffusion models suitable for generating physically consistent turbulence fields. Furthermore, to capture the dependency of turbulent attenuation with the flight speed and attitude, we extend the traditional diffusion models by incorporating the x-y spatial coordinates, speed, and attitude as conditional input~$\mathbf{C}$.
Thus, in our work, we propose the Denoising Diffusion Probabilistic Model~(DDPM) with conditional input, which consists of two main processes:

\begin{algorithm}[htp]
\caption{Diffusion Model: Training and Inference}
\label{alg:diffusion_full}
\begin{algorithmic}[1]

\Statex \textbf{Part 1: Training}
\Repeat
    \State Sample $\mathbf{X_0},\ \mathbf{C}$
    \State Sample timestep $t \sim \text{Uniform}(\{1, \ldots, T\})$
    \State Sample noise $z_t \sim \mathcal{N}(0, I)$
    \State Update parameters via gradient descent:
    \[
    \nabla_\theta \left\| z_t - \hat{z}_t^\theta \left(\sqrt{\bar{\alpha_t}} \mathbf{X_0} + \sqrt{1 - \bar{\alpha_t}} z_t, \mathbf{C},\ t \right) \right\|^2
    \]
\Until{converged}

\Statex \hrulefill
\Statex \textbf{Part 2: Inference}

\State Sample $\mathbf{X_T} \sim \mathcal{N}(0, I),\ \mathbf{C}$ 
\For{$t = T \cdots 1$}
    \If{$t > 1$}
        \State Sample $z_t \sim \mathcal{N}(0, I)$
    \Else
        \State $z_t \gets 0$
    \EndIf
    \State Compute reverse step:
    \[
    \mathbf{X_{t-1}} = \frac{1}{\sqrt{\alpha_t}} \left( \mathbf{X_t} - \frac{1 - \alpha_t}{\sqrt{1 - \alpha_t}} \epsilon_\theta(\mathbf{X_t}, \mathbf{C},\ t) \right) + \sigma_t z_t
    \]
\EndFor
\State \Return $\mathbf{X_0}$

\end{algorithmic}
\end{algorithm}

\subsubsection{Training}
During training, the raw data \( \mathbf{X_0} \) and conditional input~$\mathbf{C}$ are sampled from training dataset and defined as

\[
\mathbf{X_0} := \begin{pmatrix}
    T_0 & T_1 & \cdots & T_{N_{\textrm{batch}}-1} \\
    P_0 & P_1 & \cdots & P_{N_{\textrm{batch}}-1} \\
    B_k(\vec{r_0}) & B_k(\vec{r_0}) & \cdots & B_k(\vec{r}_{N_{\textrm{batch}}-1})
\end{pmatrix},
\]
\[
\mathbf{C} := \begin{pmatrix}
    x_1^0 & x_1^1 & \cdots & x_1^{N_{\textrm{batch}}-1} \\
    x_2^0 & x_2^1 & \cdots & x_2^{N_{\textrm{batch}}-1} \\
    M_0 & M_1 & \cdots & M_{N_{\textrm{batch}}-1} \\
    \alpha_0 & \alpha_1 & \cdots & \alpha_{N_{\textrm{batch}}-1}
\end{pmatrix},
\]
where \( N_{\textrm{batch}} \) is the batch size. Then the original data \( \mathbf{X_0} \) is corrupted step by step. The corrupted data \( \mathbf{X_t} \) at $t$ step is given as
\begin{align}
\mathbf{X_t} &= \sqrt{\bar{\alpha}_t} \mathbf{X_{0}} + \sqrt{1 - \bar{\alpha}_t} \mathbf{Z_{t}}
\label{corrupt}
\end{align}
where $\mathbf{Z_{t}}$ is sampled from standard Gaussian noise $\mathcal{N}(0,  I)$, and $\{\bar{\alpha}_t\}$ is a predefined sequence that satisfies $\lim_{t \to \infty} \bar{\alpha}_t = 0$. The training objective is to predict the added noise $\mathbf{Z_{t}}$ with given information $\mathbf{X_{0}}$ and $\mathbf{C}$, which is given as

\begin{equation}
\mathcal{L} = \left\| \mathbf{Z_{t}} - \mathbf{\hat{Z}_{t}^\theta}\left(\sqrt{\bar{\alpha}_t} \mathbf{X_0} + \sqrt{1 - \bar{\alpha}_t} \mathbf{Z_{t}}, \mathbf{C},\ t \right) \right\|^2.
\end{equation}

\subsubsection{Inference}
Based on~\eqref{corrupt}, the corrupted data at~\( t \to \infty \) step follows Standard Normal Noise:
\begin{equation}
 \mathbf{X_{t\to \infty}} \sim  \mathcal{N}(0,  \mathbf{I})
\end{equation}
Therefore, the inference process starts with $\mathbf{X_{T}} \sim  \mathcal{N}(0,  \mathbf{I})$ and iteratively denoises it to recover the original data, which is given as

\begin{align}
\mathbf{X_{t-1}} \sim \mathcal{N}(\mathbf{X_{t-1}}; \mathbf{\mu}_\theta(x_t, t), \mathbf{\Sigma}_\theta(x_t, t))
\end{align}
where \( \mathbf{\mu_\theta}(x_t, t) \) and \( \mathbf{\Sigma}_\theta(x_t, t) \) are the mean and variance predicted by the model and given as
\begin{equation}
\mu_\theta(\mathbf{X_t}, t)  =\frac{1}{\sqrt{\alpha_t}} \left(\mathbf{X_t} - \frac{1 - \alpha_t}{\sqrt{1 - \overline{\alpha}_t}} \mathbf{\hat{Z}_{t}^\theta}\right)
\end{equation}

\begin{equation}
 \Sigma(x_t, t) = \frac{(1 - \overline{\alpha}_{t-1})\cdot (1-\alpha_t)}{1 - \overline{\alpha}_t}\mathbf{I}
\end{equation}
The detailed diffusion-based surrogate model is provided in algorithm~\ref{alg:diffusion_full}.

\begin{algorithm}[t]
\caption{Iterative Power Allocation and Flight Configuration Optimization}
\label{alg:iteration}
\begin{algorithmic}[1]

\Statex \textbf{Step 1: Initialize}
\State Initialize $t \gets 0$
\State Initialize power allocation $P_{Tx}^{i,k,(0)}$ for all $k,i$
\State Initialize flight configuration $(M_k^{(0)}, \alpha_k^{(0)})$ for all $k$
\State Initialize a small value $\delta$ for convergence check
\State Initialize the initial capacity $\sum^K_{k=1} C_k^{(0)}$:
\State $C_k = \sum_{k=1}^{K} \sum_{i}\Delta f\cdot 
\log_2 \left(1 + \frac{B_k^i{P}^{k,i,(0)}_{Tx}}{L_\mathrm{turb}^{M_k^{(0)},\alpha_k^{(0)}}\cdot N}\right)$

\While{not converged}
    \State $t \gets t + 1$ 
    \State \textbf{Step 2: Solve subproblem \textbf{Q1}}
    \For{each $k, i$}
        \State $P_{Tx}^{i,k,(t)} = \left[ \frac{K \Delta f}{\lambda \ln 2} - \frac{L_\mathrm{turb}^{M_k^{(t-1)},\alpha_k^{(t-1)}}\cdot N}{A_k^i} \right]^+$
    \EndFor

    \State \textbf{Step 3: Solve subproblem \textbf{Q2}}
    \For{each $k$}
        \State $(M_k^{(t)}, \alpha_k^{(t)}) = \underset{M_k \in \mathcal{A}_M, \alpha_k \in \mathcal{A}_\alpha}{\text{argmax}} \sum_{k=1}^{K} C_k(P_{Tx}^{i,k,(t)})$
    \EndFor

    \State \textbf{Step 4: Convergence Check}
    \State Calculate the capacity $\sum^K_{k=1} C_k^{(t)}$ at iteration $t$
    \If{$|\sum^K_{k=1}C_k^{(t)} - \sum^K_{k=1}C_k^{(t-1)}| < \delta$}
        \State \textbf{Break the loop}
    \EndIf

\EndWhile

\State \Return $P_{Tx}^{i,k,(t)}$ and $(M_k^{(t)}, \alpha_k^{(t)})$

\end{algorithmic}
\end{algorithm}

After iteratively denoising to the original data $\mathbf{X_0}$, the turbulence field, including temperature, pressure, and RISC under condition $\mathbf{C}$ is quickly recovered. Thus, combining~\eqref{eq:L}, the resulting turbulent attenuation under flight speed \( M_k \) and attacking angle \( \alpha_k \) is estimated, which is essential for the accurate and efficient solution of subproblem~\textbf{Q2}. In our work, we simply employ an exhaustive search from the finite feasible sets \( \mathcal{A}_M \) and \( \mathcal{A}_\alpha \) to identify the optimal combination of flight speed \( M_k \) and attack angle \( \alpha_k \), which is given as 
\begin{equation}
(M_k^*, \alpha_k^*) = \underset{M_k \in \mathcal{A}_M, \alpha_k \in \mathcal{A}_\alpha}{\text{argmax}} \quad \sum_{k=1}^{K} C_k
\label{opt2}
\end{equation}
where \( M_k^* \) and \( \alpha_k^* \) are the optimal flight speed and attacking angle that maximize the total communication capacity.

\subsection{Iterative Solution to the Optimization Problem~\textbf{Q0}}
After solving the two subproblems~\textbf{Q1} and~\textbf{Q2}, we finally present the iterative solution to solve~\textbf{Q0}, which combines the methods of solving~\eqref{opt1} and~\eqref{opt2} to ensure both the power allocation and flight configuration are optimized simultaneously.

\it 
Theorem 2 (Iterative Solution):
The following iterative algorithm that alternates between solving the subproblems~\textbf{Q1} and~\textbf{Q2} converges to the optimal solution of the original MINLP problem.

\begin{enumerate}
\item Initialize the power allocation $P_{Tx}^{i,k,(0)}$ and the flight configuration $(M_k^{(0)},\alpha_k^{(0)})$.
\item  Solve subproblem~\textbf{Q1} given the current flight configuration $(M_k^{(t)},\alpha_k^{(t)})$ at iteration \( t \):
\begin{equation}
P_{Tx}^{i,k,(t)} = \left[ \frac{K \Delta f}{\lambda \ln 2} - \frac{L_\mathrm{turb}^{M_k^{(t)},\alpha_k^{(t)}}\cdot N}{A_k^i} \right]^+, \quad \forall k,i, \nonumber
\end{equation}
\item Solve subproblem~\textbf{Q2} given the current power allocation $P_{Tx}^{i,k,(t)}$ at iteration \( t \):
\begin{equation}
(M_k^{(t)}, \alpha_k^{(t)}) = \underset{M_k \in \mathcal{A}_M, \alpha_k \in \mathcal{A}_\alpha}{\text{argmax}} \quad \sum_{k=1}^{K} C_k(P_{Tx}^{i,k,(t)})
\nonumber
\end{equation}
\item Repeat steps 2 and 3 until convergence is achieved.
\end{enumerate}
The detailed algorithm is shown in Algorithm~\ref{alg:iteration}.

Proof: 
\rm First, the objective in~\textbf{Q0} is bounded above by the maximum achievable capacity under the power constraint \( \bar{P} \), 
\[
\sum_{k=1}^K C[k] \leq C_{\text{max}}(\bar{P}),
\]
where \( C_{\text{max}}(\bar{P}) \) is the maximum capacity achievable under the total power constraint \( \bar{P} \) and no attenuation, which is given as
\begin{align}
 C_{\text{max}}(\bar{P}) = \underset{\{ P^{k,i}_{Tx}\}}{\text{max}}  \sum_{k=1}^{K} \sum_{i}\Delta f\cdot \nonumber \log_2 \left(1 + \frac{G_{Tx}G_{Rx}P^{k,i}_{Tx}}{N}\right).  
\end{align}

Second, since we optimize the two subproblems~\textbf{Q1} and~\textbf{Q2} at each iteration, it ensures the objective value \( \sum_{k=1}^K C_k^{(t)} \) generated by the iterative algorithm is monotonically increasing: 
\[
\sum_{k=1}^K C_k^{(t+1)} \geq \sum_{k=1}^K C_k^{(t)},
\]
with equality holding only if it is already optimal.
Thus, according to the famous Bolzano-Weierstrass Theorem, 
the sequence of objective values \( \left\{ \sum_{k=1}^K C_k^{(t)} \right\} \), which is upper bounded and monotonically increasing, will converge in a finite number of steps. 

Moreover, when the sequence of objective values \( \left\{ \sum_{k=1}^K C_k^{(t)} \right\} \) is at convergence at step $T$, the iterative algorithm converges to a solution \( \left( \{ P^{k,i,(T)}_{Tx}\}, \{M_k^{(T)}, \alpha_k^{(T)}\} \right) \) that is the \textbf{joint optimal solution} for the original MINLP problem~\textbf{Q0}, due to the monotonically increasing nature of sequence $\left\{ \sum_{k=1}^K C_k^{(t)} \right\}$.
$\hfill\blacksquare$

Having proven the convergence of the iterative algorithm, we now have a comprehensive approach for solving the original optimization problem~\textbf{Q0} by decomposing it into two subproblems~\textbf{Q1} and~\textbf{Q2}, and then iteratively solving these subproblems to obtain the optimal solution under the given constraints. With the theoretical foundations established and the algorithmic steps verified, the numerical results are provided in the next section, where we analyze the practical effectiveness and efficiency of the proposed solution.

\begin{figure*}[htbp]
\centering
\subfigure[]{
\includegraphics[width=0.31\textwidth]{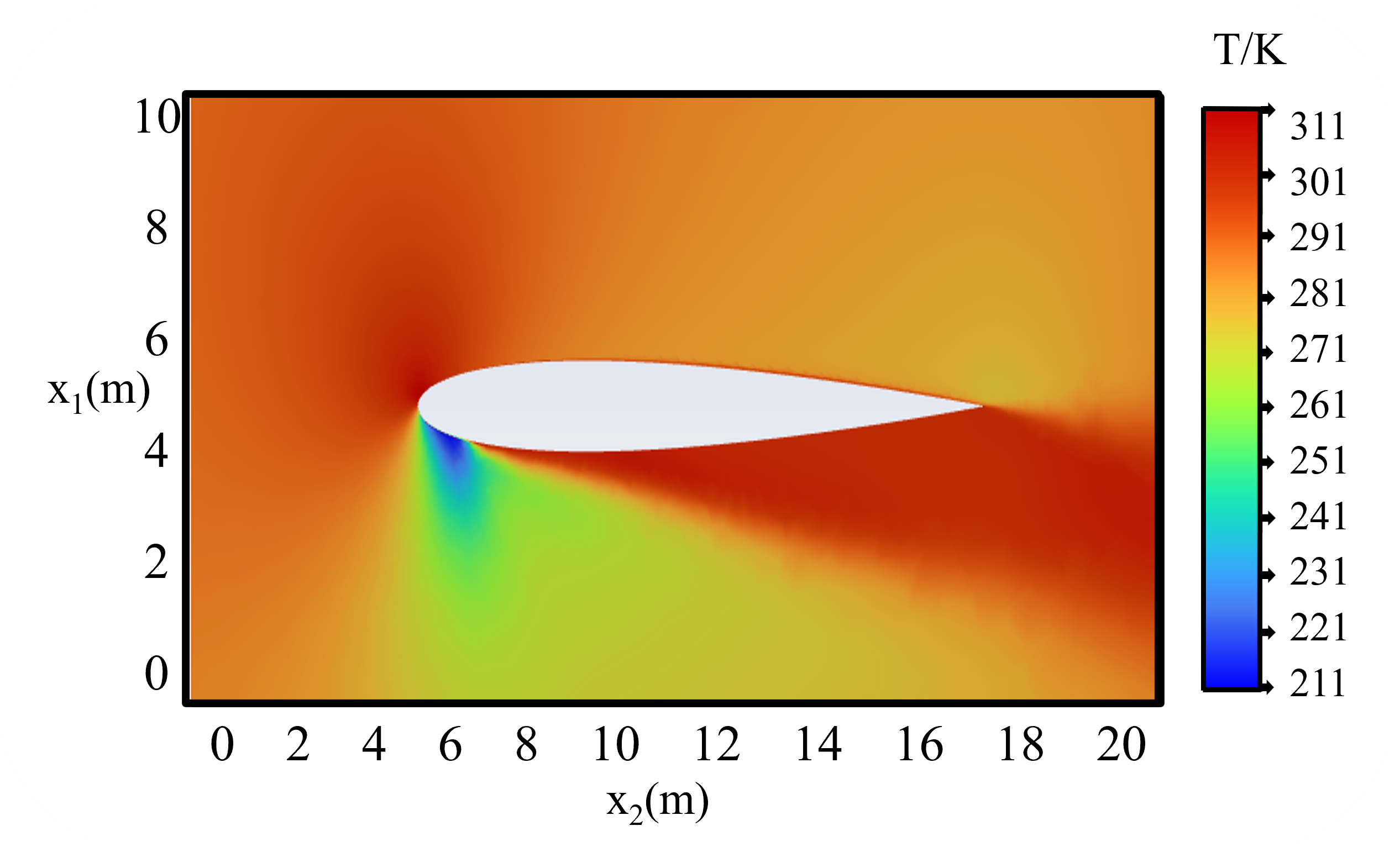}
 }
\subfigure[]{
\includegraphics[width=0.31\textwidth]{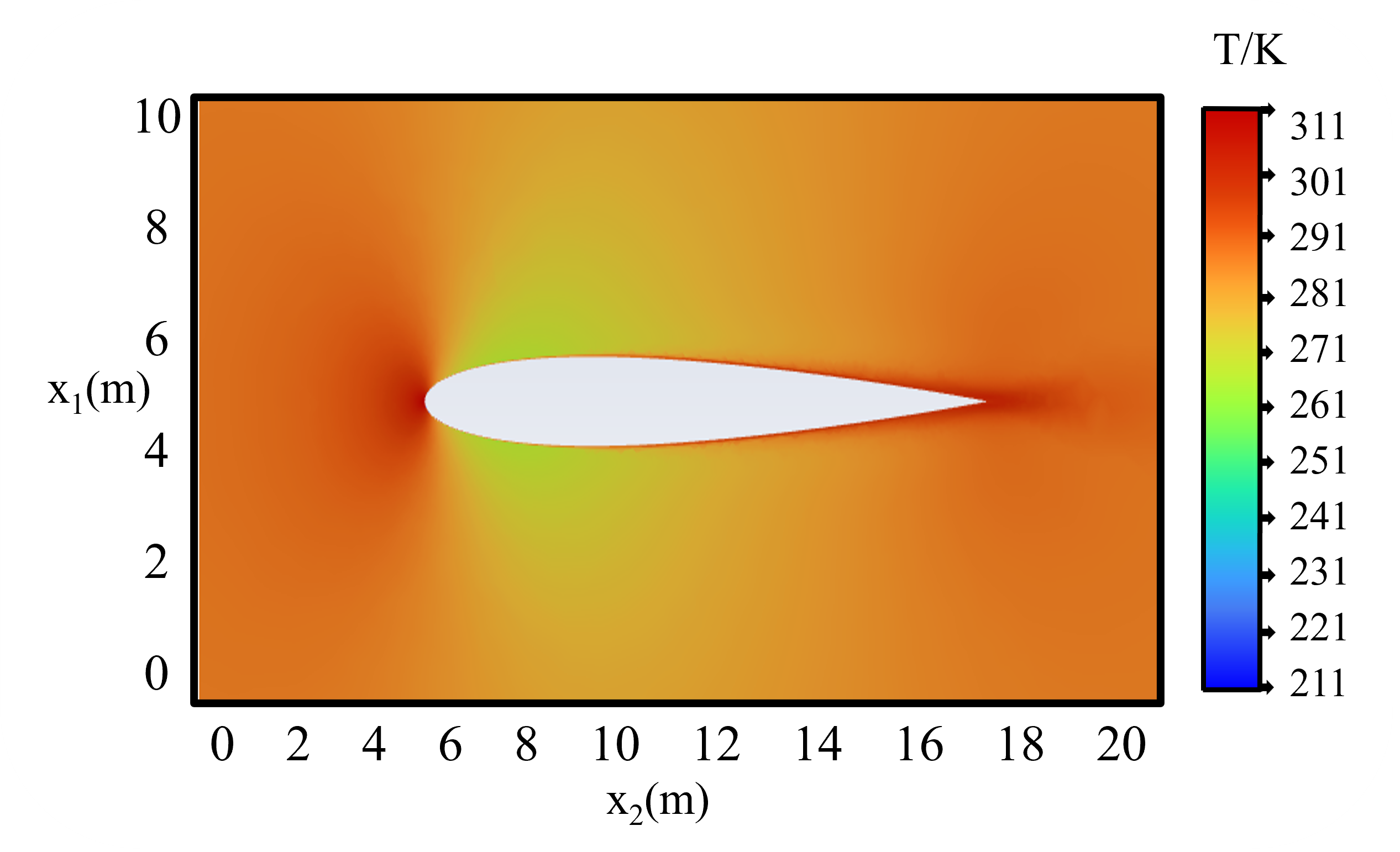}
}
\subfigure[]{
\includegraphics[width=0.31\textwidth]{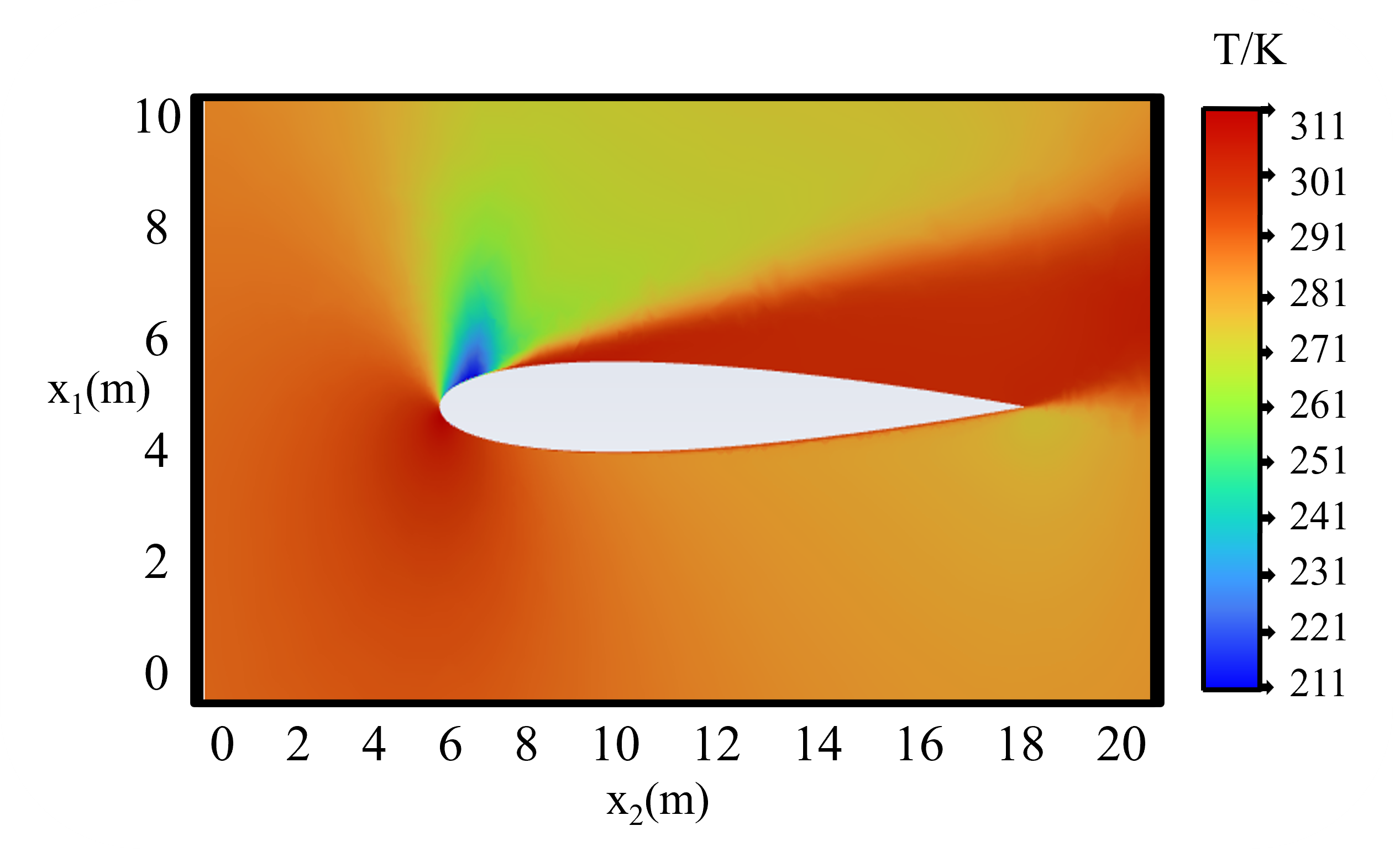}
}
\captionsetup{font={footnotesize}}
\caption{The simulated temperature field given Mach number $M = 0.7$ and different attacking angles $\alpha$. (a) $\alpha = -10^\circ$; (b) $\alpha =0^\circ$; (c) $\alpha =10^\circ$.}
\label{fig3}
\captionsetup{font={footnotesize}}
\end{figure*}

\begin{figure*}[t]
\centering
\subfigure[]{
\includegraphics[width=0.3\textwidth]{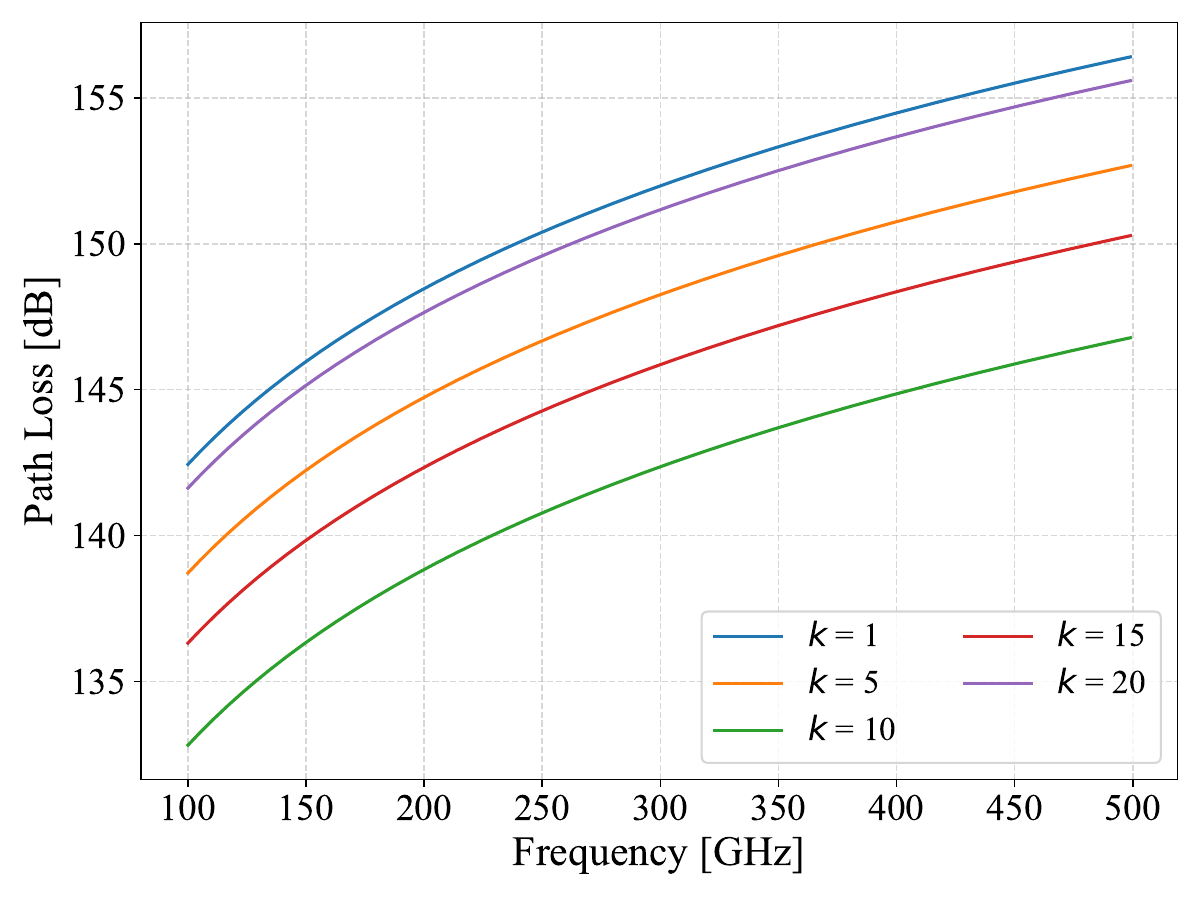}
}
\subfigure[]{
\includegraphics[width=0.3\textwidth]{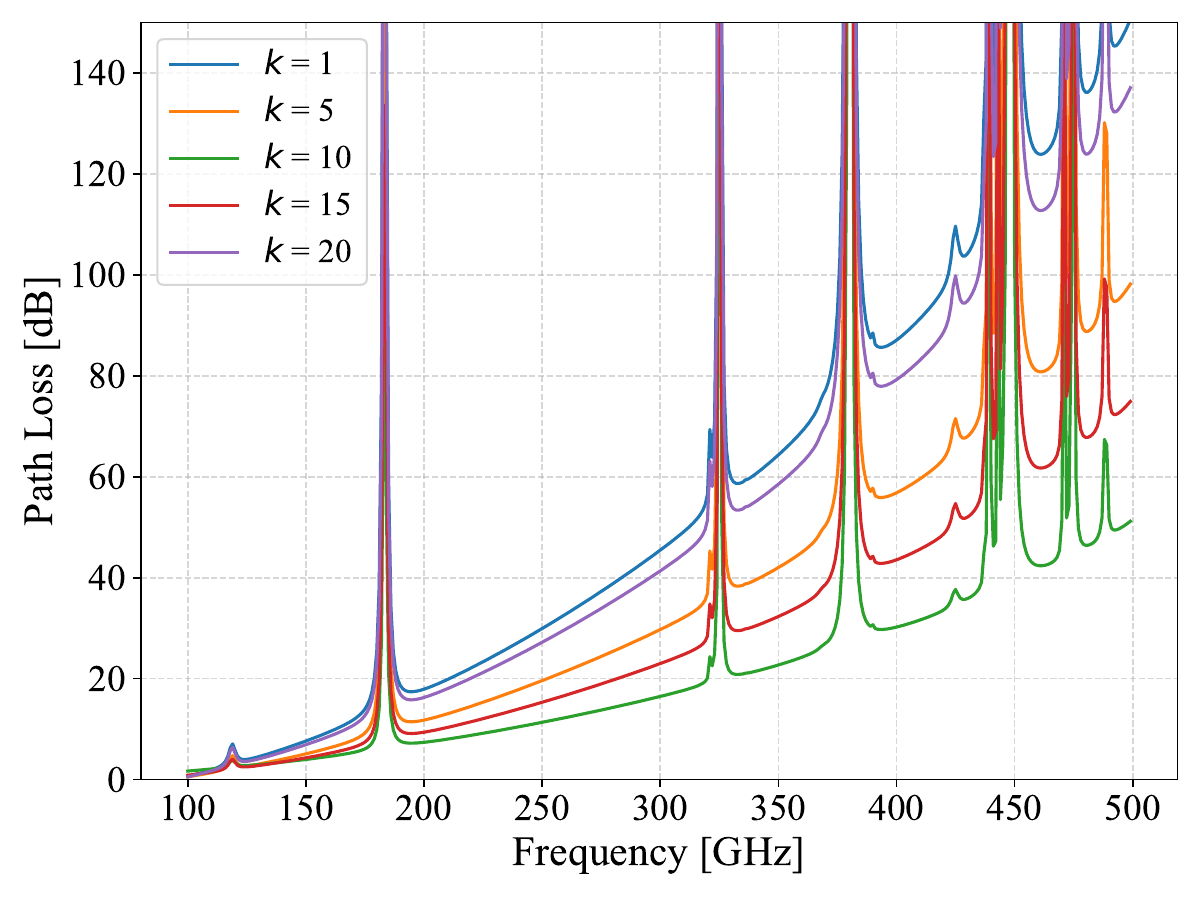}
}
\subfigure[]{
\includegraphics[width=0.3\textwidth]{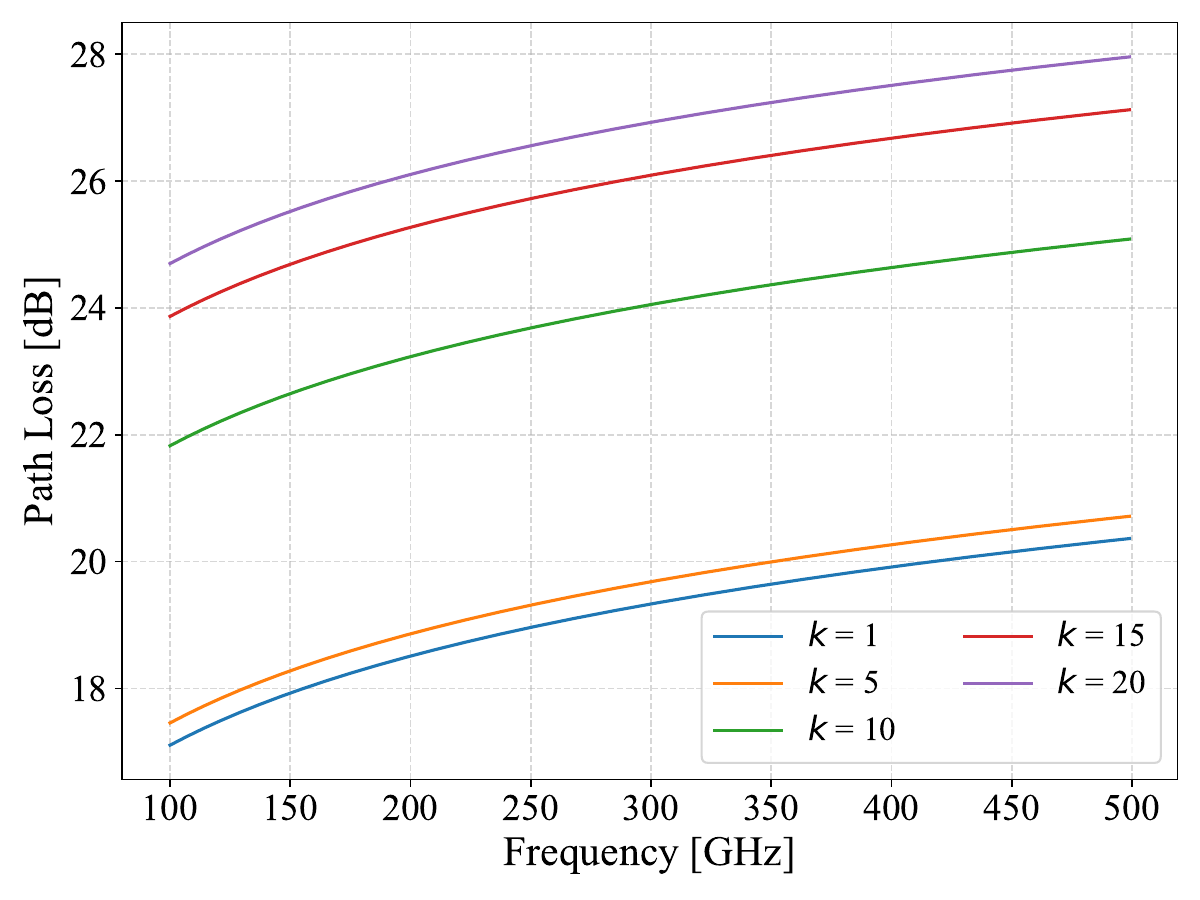}
}
\captionsetup{font={footnotesize}}
\caption{Environmental attenuation versus frequency. (a) FSPL; (b) Molecular absorption; (c) Turbulent attenuation.}
\label{fig_abs}
\captionsetup{font={footnotesize}}
\end{figure*}

\section{Numerical Results}
\label{numerical}
In this section, we numerically evaluate the proposed AI-driven optimization approach for THz communication capacity of high-speed aircraft in a turbulent environment. Specifically, the high-resolution visualizations of turbulent fields derived from the fluid dynamics-informed modeling of turbulence are first shown via simulation software Ansys Fluent. Then, the comprehensive environmental attenuation, including atmospheric water vapor absorption,
free-space path loss, and turbulence-induced attenuation, is quantified and demonstrated. Finally, the diffusion-driven optimization for THz communication
capacity is demonstrated compared with fixed flight configuration. All parameters of simulation are detailed in Table.~\ref{tab1}, unless otherwise specified.

\begin{table} 
\captionsetup{font={footnotesize}}
\renewcommand{\arraystretch}{1.3} 
\caption{Simulation Parameters}
\label{Table}
\begin{tabular}{p{1cm}p{4cm}p{1.2cm}p{0.8cm}} 
\hline  
\hline  
\textbf{Notation} & \textbf{Definition} & \textbf{Value} & \textbf{Unit}\\  
\hline 
$M$ &  flying speed & 0.7 & Mach \\ 
$\alpha$ & attacking angle & 0 & \textrm{degree} \\ 
$f$ & frequency & 100 & GHz \\ 
$H$ & flying height & 1000 & \textrm{m} \\ 
$k$ & time slot & 11 & - \\ 
$K$ & number of time slot & 21 & - \\ 
$D$ & flying distance & 6000 & \textrm{m} \\  
$\Delta f_i$ & bandwidth & 10 & \textrm{MHz} \\ 
$G_{Tx}$ & transmit antenna gain & 30 & \textrm{dBi} \\ 
$G_{Rx}$ & recieve antenna gain & 30 & \textrm{dBi} \\ 
$N_0$ & noise power spectrum & -169 & \textrm{dBm/Hz} \\ 
$\bar{P}$ & average power & 10 & \textrm{dBm} \\ 
$\bar{M}$ & average Mach number & 0.6 & - \\ 
$\mathcal{A}_M$ & feasible domain of Mach number & 0.5, 0.7 & - \\ 
$\mathcal{A}_\alpha$ & feasible domain of attacking angles & 0, -10, 10 & \textrm{degree} \\ 
\hline
\hline  
\end{tabular}  
\label{tab1}
\end{table}

\begin{table*}[ht]  
\captionsetup{font={footnotesize}}
\caption{Average Channel Capacities for Optimized and Average Power Allocation under different flight configurations.} 
\label{table:fading}
\centering
\renewcommand\arraystretch{1.3} % 调大行距
\begin{tabular}{p{3cm}p{3cm}p{3cm}p{3cm}p{3cm}} 
\hline  
\hline  
\textbf{Mach Number} & \textbf{Attacking Angle} & \textbf{Capacity $\bar{C_0}$} & \textbf{Capacity $\bar{C^*}$} & \textbf{Improvement}\\ 
\hline 
$M=0.5$ & $\alpha = 0^\circ$ & $10.643~\textrm{bps}/\textrm{Hz}$ & $14.318~\textrm{bps}/\textrm{Hz}$ & 34.5\%\\
$M=0.5$ & $\alpha = 10^\circ$ & $14.118~\textrm{bps}/\textrm{Hz}$ & $18.135~\textrm{bps}/\textrm{Hz}$ & 28.5\%\\
$M=0.5$ & $\alpha = -10^\circ$ & $11.793~\textrm{bps}/\textrm{Hz}$ & $15.578~\textrm{bps}/\textrm{Hz}$ & 32.1\%\\
$M=0.7$ & $\alpha = 0^\circ$ & $6.753~\textrm{bps}/\textrm{Hz}$ & $9.881~\textrm{bps}/\textrm{Hz}$ & 46.3\%\\
$M=0.7$ & $\alpha = 10^\circ$ & $8.981~\textrm{bps}/\textrm{Hz}$ & $12.462~\textrm{bps}/\textrm{Hz}$ & 38.8\%\\
$M=0.7$ & $\alpha = -10^\circ$ & $6.857~\textrm{bps}/\textrm{Hz}$ & $9.972~\textrm{bps}/\textrm{Hz}$ & 45.4\%\\
\hline
\hline  
\end{tabular}  
\label{tab_Q2}
\end{table*}

\begin{figure}[t]
\centerline{\includegraphics[width=0.5\textwidth]{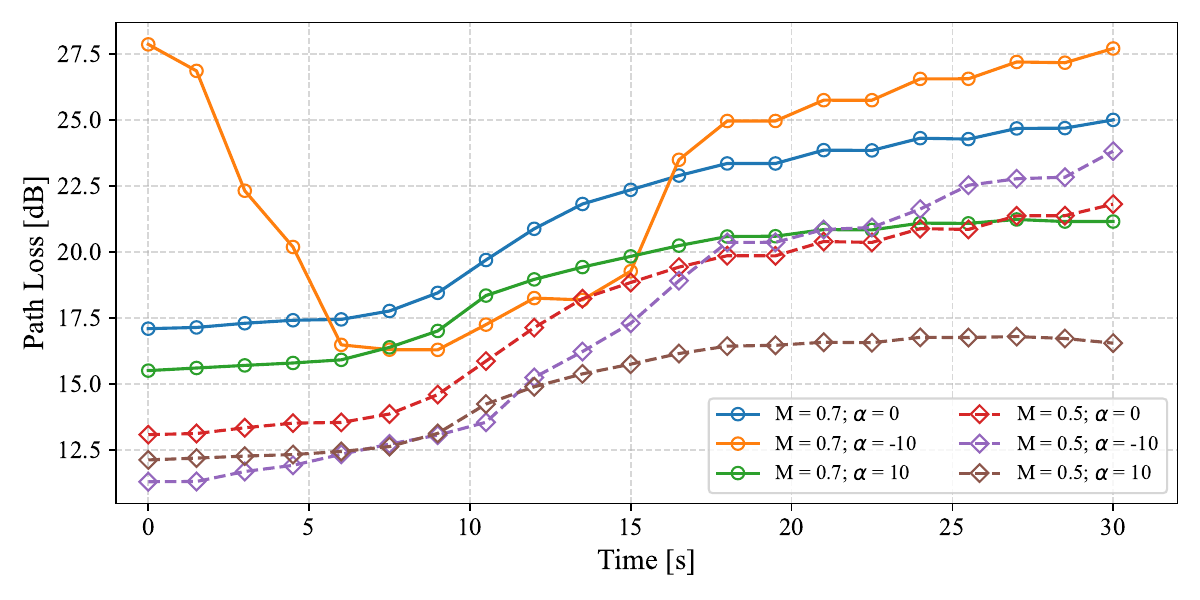}} 
\captionsetup{font={footnotesize}}
\caption{Turbulent attenuation over times at different $M$ and $\alpha$.}
\label{turb_k}
\end{figure}

\subsection{Fluid Dynamic-informed Computation and Analysis of Turbulent Field}
Based on the governing equations of turbulence fields in the RANS approach and SST model~\eqref{eq6}-\eqref{eq_k2}, the turbulence fields including temperature $T$, pressure $P$, turbulent kinetic energy $\mathcal{E}$ and energy dissipation rate $\omega$ around the high-speed aircraft under different boundary conditions~$\left(M, \alpha\right)$ can be obtained by solving~\eqref{eq6}-\eqref{eq_k2}. For instance, the spatial distributions of temperature are shown in Fig.~\ref{fig3} under $M = 0.7$ and $\alpha =0^\circ,\pm10^\circ$. As we can observe, the spatial distribution of temperature near the aircraft surface is highly uneven and exhibits distinct deviations compared to the temperature further from the aircraft. This indicates that the turbulent flow around the high-speed aircraft plays a critical role in creating non-uniform distributions of temperature fields on the surface. Besides, compared between Fig.~\ref{fig3}(a), Fig.~\ref{fig3}(b) and Fig.~\ref{fig3}(c), we observe that variations in $\alpha$ further influence the temperature distribution. Specifically, at $\alpha = \pm 10^\circ$,  the distributions of temperature are more heterogeneous compared with the distribution of temperature at $\alpha = 0^\circ$, which implies  that higher magnitude of $\alpha$ leads to more irregular temperature fields.  Similar trends are observed in the spatial distributions of pressure $P$, turbulent kinetic energy $\mathcal{E}$ and energy dissipation rate $\omega$.
These observations align with fundamental aerodynamics principles and demonstrate the increasing inhomogeneity of the turbulent fields with the growing magnitude of $
\alpha$.

\subsection{Comprehensive Environmental Attenuation in THz AG Channel}
Based on~\cite{jornet2011channel, series2019attenuation} and the fluid dynamic-informed modeling of turbulence, the comprehensive environmental attenuation, including molecular absorption, free-space path loss, and turbulence-induced attenuation, can be quantified. Fig.~\ref{fig_abs} demonstrates the three kinds of attenuation over frequency band $100-500~\textrm{GHz}$ at $H = 1$ km and different time slots. As we can observe, FSPL and molecular absorption remain the dominant parts of attenuation in THz channel. Meanwhile, the impact of turbulence-induced attenuation, ranging from $18-28~\textrm{dB}$, is also non-negligible, particularly when assessing realistic AG communication in our scenarios. In addition, it can be observed that all these three types of attenuation exhibit strong frequency-selective characteristics, especially molecular absorption, where strong absorption peaks occur in specific sub-bands, resulting in severe performance degradation. Consequently, it is essential to incorporate frequency-selective modeling in the channel capacity in~\eqref{capa}, to accurately capture the effects of the three kinds of attenuation. 

Furthermore, Fig.~\ref{turb_k} demonstrates the vertical distribution of the turbulence-induced attenuation over time slots $k$ and varying Mach numbers $M$ and attacking angles~$\alpha$. It can be observed that the attenuation at $M=0.7$ is around $22~\textrm{dB}$, which is generally higher than the attenuation at $M=0.5$ with an average of $18~\textrm{dB}$. This suggests that higher aircraft speeds $M$ lead to stronger turbulent flow and the resulting attenuation, consistent with the earlier conclusion that the turbulence intensity increases with speed. However, the attenuation at the same $M$ and different $\alpha$ fluctuates substantially along the time slots and no single attacking angle $\alpha$ maintains minimal attenuation across all slots, which justifies the need for real-time attitude optimization to maintain reliable communication links. 

\begin{figure}[t]
\centerline{\includegraphics[width=0.5\textwidth]{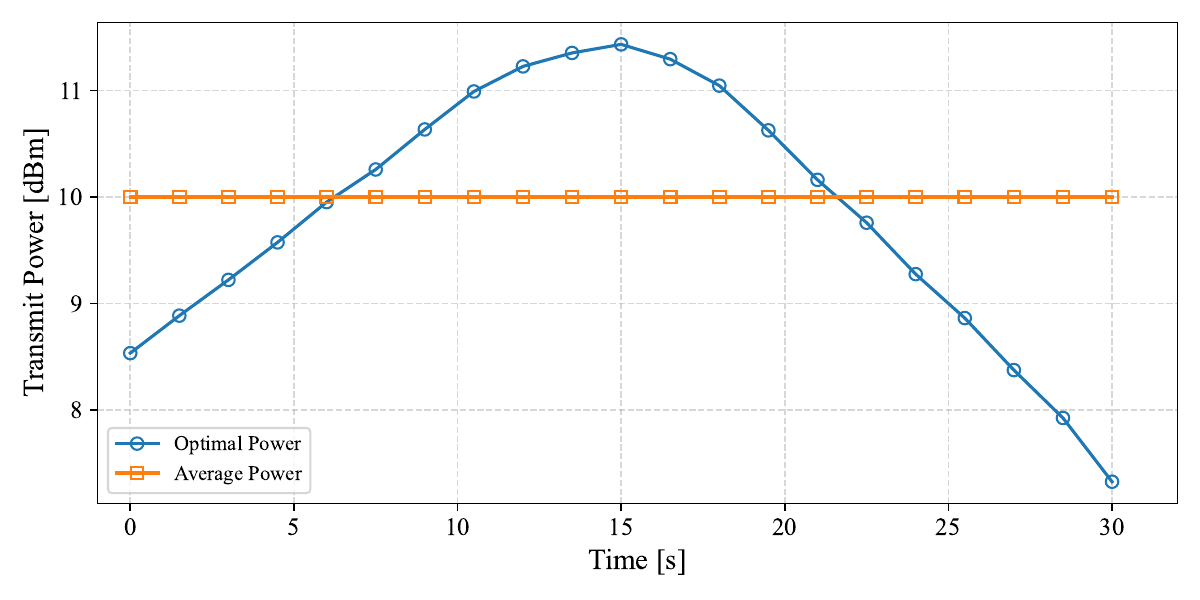}} 
\captionsetup{font={footnotesize}}
\caption{Optimized and average power allocation over times.}
\label{fig_Q1}
\end{figure}

\subsection{Diffusion-Based Optimization of Capacity via Iterative solution}
We optimize the THz communication
capacity for high-speed aircraft in AG channel in Fig.~\ref{fig_environment} by invoking diffusion-based surrogate model and iterative solution. Since the optimization problem~\textbf{Q0} is decomposed into two subproblem~\textbf{Q1} and~\textbf{Q2}, we first demonstrate the optimization results of~\textbf{Q1} and~\textbf{Q2}.
\subsubsection{Optimization of~\textbf{Q1}}
Based on the generalized water-filling approach in Theorem 1, the result of power optimization with fixed flight configuration is demonstrated in Fig.~\ref{fig_Q1}. As we can observe, the optimized allocation of transmit power over time slots exhibits a distinct temporal pattern compared with the average power allocation. Specifically, the optimized scheme allocates relatively higher power levels to the middle time slots while assigning lower power levels to both ends. This behavior indicates that the optimization process effectively concentrates transmission resources in the time periods where the channel attenuation is smaller and channel conditions are more favorable, thereby improving the overall system capacity.
Furthermore, the average channel capacities for both the optimized power allocation and the average power allocation, denoted as $C^*$ and $C_0$, are presented in Table.~\ref{tab_Q2} under different flight configurations. As we can observe, the optimized algorithm consistently outperforms the average power allocation strategy, leading to an average channel capacity improvement of approximately 10\%. Moreover, when under the same attacking angle $\alpha$, the improvement achieved at $M=0.7$ is consistently larger than that at $M=0.5$, which demonstrates the advantage of the optimized scheme in scenarios with more challenging flight dynamics.

\begin{figure}[t]
\centerline{\includegraphics[width=0.5\textwidth]{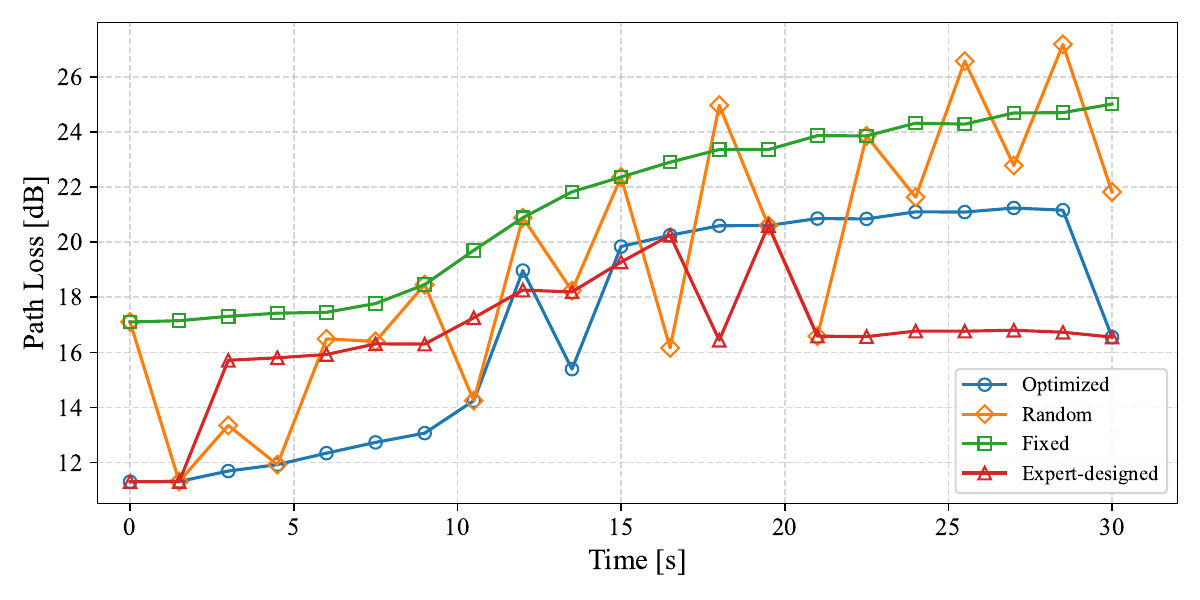}} 
\captionsetup{font={footnotesize}}
\caption{Turbulent attenuation over time slots under different strategies.}
\label{L_Q2}
\end{figure}

\begin{table}[t]
\captionsetup{font={footnotesize}}
\renewcommand{\arraystretch}{1.3}
\caption{Average turbulent attenuation and channel capacities for different strategies of flight configurations.}
\label{Table}
\begin{tabular}{p{2.5cm}p{2.5cm}p{2.5cm}} 
\hline  
\hline  
\textbf{Strategy} & \textbf{Attenuation ${\bar{L}_\mathrm{turb}}$} & \textbf{Capacity $\bar{C}$}\\ 
\hline 
Expert-designed & $19.02~\textrm{dB}$ & $11.384~\textrm{bps}/\textrm{Hz}$\\ 
Optimized    & $21.81~\textrm{dB}$ & $11.241~\textrm{bps}/\textrm{Hz}$\\ 
Random     & $28.33~\textrm{dB}$ & $9.147~\textrm{bps}/\textrm{Hz}$\\ 
Fixed   & $18.66~\textrm{dB}$ & $6.752~\textrm{bps}/\textrm{Hz}$ \\ 
\hline
\hline  
\end{tabular}  
\label{tab_Q2}
\end{table}

\subsubsection{Optimization of~\textbf{Q2}}
Given the diffusion-based surrogate model, the turbulence-induced attenuation under different $M$ and $\alpha$ can be estimated almost instantly. Hence, \textbf{Q2} can be optimized by exhaustively searching the optimized sequence $\{M_k^*, \alpha_k^*\}$ that minimizes overall $L_\mathrm{turb}^{M_k,\alpha_k}$ and thus maximizes the THz capacity in \eqref{sub2}. 
Fig.~\ref{L_Q2} demonstrates the turbulence-induced attenuation $L_\mathrm{turb}^{M_k,\alpha_k}$ under $\{M_k^*, \alpha_k^*\}$ compared with three baseline strategies, namely the expert-designed sequence, a randomly generated sequence and a fixed sequence. In particular, the expert sequence is designed based on prior full knowledge of turbulence, and the fixed strategy corresponds to maintaining $M=0.7$ and $\alpha=0^\circ$ throughout the entire flight. Moreover, the average turbulence-induced attenuation and the average capacity, denoted as $\bar{L}_\mathrm{turb}$ and $\bar{C}$, are shown in Table.~\ref{tab_Q2}. It is worth noting that the transmit power allocation is kept fixed in the optimization problem \textbf{Q2}, and the average power allocation scheme is adopted just to simplify the calculation of $\bar{C}$. As we can observe, the optimized sequence reduces $\bar{L}_\mathrm{turb}$ and achieves the capacity improvements of about 22.8\% and 66.5\% compared with random strategy and fixed strategy, respectively. In addition, the expert sequence slightly outperforms the optimized strategy, which suggests that the proposed diffusion-aided optimization is able to approach expert-level performance, while substantially outperforming other baselines.

\begin{figure}[t]
\centerline{\includegraphics[width=0.5\textwidth]{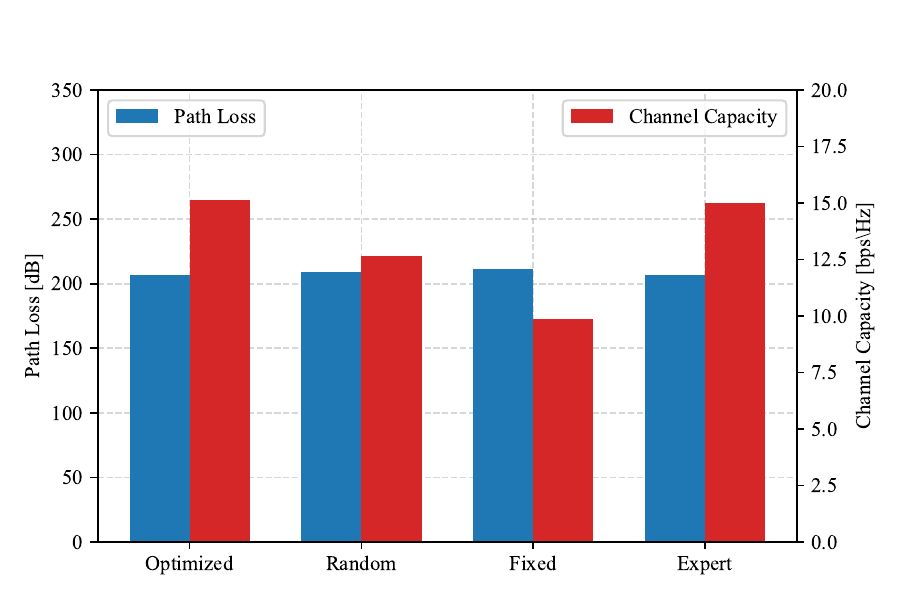}} 
\captionsetup{font={footnotesize}}
\caption{Molecular absorption over frequency at different time slots.}
\label{Q_final}
\end{figure}

Finally, by alternating between the subproblems \textbf{Q1} and \textbf{Q2} within an iterative optimization loop, the high-quality solution to the original joint problem \textbf{Q0} can be obtained. As demonstrated in Fig.~\ref{Q_final}, we compare the average of comprehensive environmental attenuation and capacity under four different flight configuration strategies and optimized allocation of transmit power. It is observed that the average capacity reaches up to $11.241~\textrm{bps}/\textrm{Hz}$ under our optimized approach, while the random strategy and fixed strategy remain below $10~\textrm{bps}/\textrm{Hz}$. Moreover, our optimized strategy approaches the expert results very closely, with only a marginal gap in capacity. These results highlight the clear advantage of jointly optimizing flight configuration and power allocation to solve \textbf{Q0} effectively.

\section{Conclusion}
This paper proposes an AI-empowered THz AG communication framework to accurately capture the dynamic turbulence-induced attenuation driven by variations in aircraft speed and attitude. A fluid dynamics-informed system model characterizes turbulence effects on THz links, and a joint power-attitude optimization framework maximizes communication capacity. To reduce computational complexity, a diffusion-based surrogate model efficiently estimates the additional loss and supports the iterative algorithm. Numerical results demonstrate that turbulence-induced attenuation ranges from $18~\textrm{dB}$ to $28~\textrm{dB}$, and varies significantly with aircraft speed and attitude, highlighting the critical impact of aircraft-induced turbulence on THz AG links. 
Furthermore, the proposed framework achieves an average capacity of up to $11.241~\textrm{bps}/\textrm{Hz}$, outperforming random and fixed strategies that remain below $10~\textrm{bps}/\textrm{Hz}$, and closely approaching expert benchmarks.
These findings underscore the importance of integrating fluid dynamics-informed turbulence modeling with AI-based optimization to enable reliable, high-capacity THz communications in future SAGINs.

\label{conclusion}

\bibliographystyle{ieeetr}
\bibliography{ref}

@article{yang20196g,
  title={6{G} wireless communications: Vision and potential techniques},
  author={Yang, Ping and Xiao, Yue and Xiao, Ming and Li, Shaoqian},
  journal={IEEE {N}etwork},
  volume={33},
  number={4},
  pages={70--75},
  year={2019},
  publisher={IEEE}
}

@inproceedings{stotts2022bit,
  title={Bit error rate performance of a laser ground-to-satellite uplink communications systems in the presence of atmospheric turbulence and loss},
  author={Stotts, Larry B and Andrews, Larry C},
  booktitle={2022 IEEE International Conference on Space Optical Systems and Applications (ICSOS)},
  pages={66--73},
  year={2022},
  organization={IEEE}
}

@article{gao2025seamless,
  title={Seamless Gbps Hybrid FSO/THz Communication With Intelligent Switching},
  author={Gao, Weijun and Liu, Ke and Han, Chong and Chen, Zhi},
  journal={Journal of Lightwave Technology},
  year={2025},
  publisher={IEEE}
}

@article{cang2019impact,
  title={The impact of atmospheric turbulence on {T}erahertz communication},
  author={Cang, Lei and Zhao, Heng-Kai and Zheng, Guo-Xin},
  journal={IEEE Access},
  volume={7},
  pages={88685--88692},
  year={2019},
  publisher={IEEE}
}

@article{al2001mathematical,
  title={Mathematical model for the irradiance probability density function of a laser beam propagating through turbulent media},
  author={Al-Habash, MA and Andrews, Larry C and Phillips, Ronald L},
  journal={Optical {E}ngineering},
  volume={40},
  number={8},
  pages={1554--1562},
  year={2001},
  publisher={Society of Photo-Optical Instrumentation Engineers}
}

@article{dordova2010calculation,
  title={Calculation and comparison of turbulence attenuation by different methods},
  author={Dordov{\'a}, Lucie and Wilfert, Otakar},
  journal={Radio {E}ngineering},
  volume={19},
  number={1},
  pages={162--167},
  year={2010}
}

@article{david20186g,
  title={6{G} vision and requirements: Is there any need for beyond 5{G}?},
  author={David, Klaus and Berndt, Hendrik},
  journal={IEEE Vehicular Technology Magazine},
  volume={13},
  number={3},
  pages={72--80},
  year={2018},
  publisher={IEEE}
}

@article{cui2019measurement,
  title={Measurement-based modeling and analysis of {UAV} air-ground channels at 1 and 4 {GH}z},
  author={Cui, Zhuangzhuang and Briso-Rodr{\'\i}guez, C{\'e}sar and Guan, Ke and Calvo-Ramirez, Cesar and Ai, Bo and Zhong, Zhangdui},
  journal={IEEE Antennas and Wireless Propagation Letters},
  volume={18},
  number={9},
  pages={1804--1808},
  year={2019},
  publisher={IEEE}
}

@article{zeng2016wireless,
  title={Wireless communications with unmanned aerial vehicles: Opportunities and challenges},
  author={Zeng, Yong and Zhang, Rui and Lim, Teng Joon},
  journal={IEEE Communications {M}agazine},
  volume={54},
  number={5},
  pages={36--42},
  year={2016},
  publisher={IEEE}
}

@article{liu2021uav,
  title={An {UAV}-enabled intelligent connected transportation system with 6{G} communications for Internet of Vehicles},
  author={Liu, Run and Liu, Anfeng and Qu, Zhenzhe and Xiong, Neal N},
  journal={IEEE Transactions on Intelligent Transportation Systems},
  volume={24},
  number={2},
  pages={2045--2059},
  year={2021},
  publisher={IEEE}
}

@article{liu2021novel,
  title={A novel non-stationary 6{G} {UAV} channel model for maritime communications},
  author={Liu, Yu and Wang, Cheng-Xiang and Chang, Hengtai and He, Yubei and Bian, Ji},
  journal={IEEE Journal on Selected Areas in Communications},
  volume={39},
  number={10},
  pages={2992--3005},
  year={2021},
  publisher={IEEE}
}

@article{petrov2020ieee,
  title={{IEEE} 802.15. 3d: First standardization efforts for sub-{T}erahertz band communications toward 6{G}},
  author={Petrov, Vitaly and Kurner, Thomas and Hosako, Iwao},
  journal={IEEE Communications Magazine},
  volume={58},
  number={11},
  pages={28--33},
  year={2020},
  publisher={IEEE}
}

@article{chen2021terahertz,
  title={{T}erahertz wireless communications for 2030 and beyond: A cutting-edge frontier},
  author={Chen, Zhi and Han, Chong and Wu, Yongzhi and Li, Lingxiang and Huang, Chongwen and Zhang, Zhaoyang and Wang, Guangjian and Tong, Wen},
  journal={IEEE Communications Magazine},
  volume={59},
  number={11},
  pages={66--72},
  year={2021},
  publisher={IEEE}
}

@article{jiang2024terahertz,
  title={{T}erahertz communications and sensing for 6{G} and beyond: A comprehensive review},
  author={Jiang, Wei and Zhou, Qiuheng and He, Jiguang and Habibi, Mohammad Asif and Melnyk, Sergiy and El-Absi, Mohammed and Han, Bin and Di Renzo, Marco and Schotten, Hans Dieter and Luo, Fa-Long},
  journal={IEEE Communications Surveys \& Tutorials},
  volume={26},
  number={4},
  pages={2326--2381},
  year={2024},
  publisher={IEEE}
}

@article{huang2024energy,
  title={Energy efficiency maximization in {UAV}-assisted intelligent autonomous transport system for 6{G} networks with energy harvesting},
  author={Huang, Jie and Yu, Tao and Zhu, Xiaogang and Yang, Fan and Lai, Xianzhi and Alfarraj, Osama and Yu, Keping},
  journal={IEEE Transactions on Intelligent Transportation Systems},
  year={2024},
  volume={},
  number={},
  pages={1-11},
  publisher={IEEE}
}

@article{geraci2022will,
  title={What will the future of {UAV} cellular communications be? A flight from 5{G} to 6{G}},
  author={Geraci, Giovanni and Garcia-Rodriguez, Adrian and Azari, M Mahdi and Lozano, Angel and Mezzavilla, Marco and Chatzinotas, Symeon and Chen, Yun and Rangan, Sundeep and Di Renzo, Marco},
  journal={IEEE communications Surveys \& Tutorials},
  volume={24},
  number={3},
  pages={1304--1335},
  year={2022},
  publisher={IEEE}
}

@article{xiao2024space,
  title={Space-air-ground integrated wireless networks for 6{G}: Basics, key technologies and future trends},
  author={Xiao, Yue and Ye, Ziqiang and Wu, Mingming and Li, Haoyun and Xiao, Ming and Alouini, Mohamed-Slim and Al-Hourani, Akram and Cioni, Stefano},
  journal={IEEE Journal on Selected Areas in Communications},
  year={2024},
  publisher={IEEE}
}

@article{cui2022space,
  title={Space-air-ground integrated network ({SAGIN}) for 6{G}: Requirements, architecture and challenges},
  author={Cui, Huanxi and Zhang, Jun and Geng, Yuhui and Xiao, Zhenyu and Sun, Tao and Zhang, Ning and Liu, Jiajia and Wu, Qihui and Cao, Xianbin},
  journal={China Communications},
  volume={19},
  number={2},
  pages={90--108},
  year={2022},
  publisher={IEEE}
}

@article{dang2020should,
  title={What should 6{G} be?},
  author={Dang, Shuping and Amin, Osama and Shihada, Basem and Alouini, Mohamed-Slim},
  journal={Nature Electronics},
  volume={3},
  number={1},
  pages={20--29},
  year={2020},
  publisher={Nature Publishing Group UK London}
}

@article{zhou2021gateway,
  title={Gateway placement in integrated satellite--terrestrial networks: Supporting communications and Internet of Remote Things},
  author={Zhou, Di and Sheng, Min and Wu, Jiaxin and Li, Jiandong and Han, Zhu},
  journal={IEEE Internet of Things Journal},
  volume={9},
  number={6},
  pages={4421--4434},
  year={2021},
  publisher={IEEE}
}

@article{zhao2019uav,
  title={{UAV}-assisted emergency networks in disasters},
  author={Zhao, Nan and Lu, Weidang and Sheng, Min and Chen, Yunfei and Tang, Jie and Yu, F Richard and Wong, Kai-Kit},
  journal={IEEE Wireless Communications},
  volume={26},
  number={1},
  pages={45--51},
  year={2019},
  publisher={IEEE}
}

@article{chen2020securing,
  title={Securing aerial-ground transmission for {NOMA-UAV} networks},
  author={Chen, Xinying and Li, Dongdong and Yang, Zhutian and Chen, Yunfei and Zhao, Nan and Ding, Zhiguo and Yu, F Richard},
  journal={IEEE Network},
  volume={34},
  number={6},
  pages={171--177},
  year={2020},
  publisher={IEEE}
}

@article{zhu2020millimeter,
  title={Millimeter-wave full-duplex {UAV} relay: Joint positioning, beamforming, and power control},
  author={Zhu, Lipeng and Zhang, Jun and Xiao, Zhenyu and Cao, Xianbin and Xia, Xiang-Gen and Schober, Robert},
  journal={IEEE Journal on Selected Areas in Communications},
  volume={38},
  number={9},
  pages={2057--2073},
  year={2020},
  publisher={IEEE}
}

@article{saad2019vision,
  title={A vision of 6{G} wireless systems: Applications, trends, technologies, and open research problems},
  author={Saad, Walid and Bennis, Mehdi and Chen, Mingzhe},
  journal={IEEE {N}etwork},
  volume={34},
  number={3},
  pages={134--142},
  year={2019},
  publisher={IEEE}
}

@article{han2022terahertz,
  title={{T}erahertz wireless channels: A holistic survey on measurement, modeling, and analysis},
  author={Han, Chong and Wang, Yiqin and Li, Yuanbo and Chen, Yi and Abbasi, Naveed A and K{\"u}rner, Thomas and Molisch, Andreas F},
  journal={IEEE Communications Surveys \& Tutorials},
  volume={24},
  number={3},
  pages={1670--1707},
  year={2022},
  publisher={IEEE}
}

@inproceedings{li2021ray,
  title={Ray-tracing simulation and hybrid channel modeling for low-{T}erahertz {UAV} communications},
  author={Li, Yuanbo and Li, Ning and Han, Chong},
  booktitle={ICC 2021-IEEE International Conference on Communications},
  pages={1--6},
  year={2021},
  organization={IEEE}
}

@article{kokkoniemi2021channel,
  title={Channel modeling and performance analysis of airplane-satellite {T}erahertz band communications},
  author={Kokkoniemi, Joonas and Jornet, Josep M and Petrov, Vitaly and Koucheryavy, Yevgeni and Juntti, Markku},
  journal={IEEE Transactions on Vehicular Technology},
  volume={70},
  number={3},
  pages={2047--2061},
  year={2021},
  publisher={IEEE}
}

@article{series2019attenuation,
  title={Attenuation by atmospheric gases and related effects},
  author={Series, P},
  journal={Recommendation ITU-R},
  volume={25},
  pages={676--12},
  year={2019}
}

@article{jornet2011channel,
  title={Channel modeling and capacity analysis for electromagnetic wireless nanonetworks in the {T}erahertz band},
  author={Jornet, Josep Miquel and Akyildiz, Ian F},
  journal={IEEE Transactions on Wireless Communications},
  volume={10},
  number={10},
  pages={3211--3221},
  year={2011},
  publisher={IEEE}
}

@inproceedings{an2022measurement,
  title={Measurement and ray-tracing for {UAV} air-to-air channel modeling},
  author={An, Hao and Guan, Ke and Li, Wenbin and Zhang, Jundi and He, Danping and Zhu, Fusheng and Chen, Lei},
  booktitle={2022 IEEE 5th International Conference on Electronic Information and Communication Technology (ICEICT)},
  pages={415--420},
  year={2022},
  organization={IEEE}
}

@article{cui2022cluster,
  title={Cluster-based characterization and modeling for {UAV} air-to-ground time-varying channels},
  author={Cui, Zhuangzhuang and Guan, Ke and Oestges, Claude and Briso-Rodr{\'\i}guez, C{\'e}sar and Ai, Bo and Zhong, Zhangdui},
  journal={IEEE Transactions on Vehicular Technology},
  volume={71},
  number={7},
  pages={6872--6883},
  year={2022},
  publisher={IEEE}
}

@article{mao2022terahertz,
  title={{T}erahertz-band near-space communications: From a physical-layer perspective},
  author={Mao, Tianqi and Zhang, Leyi and Xiao, Zhenyu and Han, Zhu and Xia, Xiang-Gen},
  journal={IEEE Communications {M}agazine},
  volume={62},
  number={2},
  pages={110--116},
  year={2022},
  publisher={IEEE}
}

@article{zhou2023aerospace,
  title={Aerospace integrated networks innovation for empowering 6{G}: A survey and future challenges},
  author={Zhou, Di and Sheng, Min and Li, Jiandong and Han, Zhu},
  journal={IEEE Communications Surveys \& Tutorials},
  volume={25},
  number={2},
  pages={975--1019},
  year={2023},
  publisher={IEEE}
}

@article{wang20226g,
  title={6{G} {TH}z propagation channel characteristics and modeling: Recent developments and future challenges},
  author={Wang, Jun and Wang, Cheng-Xiang and Huang, Jie and Chen, Yunfei},
  journal={IEEE Communications Magazine},
  volume={62},
  number={2},
  pages={56--62},
  year={2022},
  publisher={IEEE}
}

@ARTICLE{gao2024attenuation,
  author={Gao, Weijun and Han, Chong and Chen, Zhi},
  journal={IEEE Transactions on Wireless Communications}, 
  title={Attenuation and Loss of Spatial Coherence Modeling for Atmospheric Turbulence in {T}erahertz {UAV} {MIMO} Channels}, 
  year={2024},
  volume={23},
  number={9},
  pages={11636-11648},
  doi={10.1109/TWC.2024.3384036}}

@article{ma2015experimental,
  title={Experimental comparison of {T}erahertz and infrared signaling in controlled atmospheric turbulence},
  author={Ma, Jianjun and Moeller, Lothar and Federici, John F},
  journal={Journal of Infrared, Millimeter, and {T}erahertz Waves},
  volume={36},
  pages={130--143},
  year={2015},
  publisher={Springer}
}

@book{tatarski2016wave,
  title={Wave propagation in a turbulent medium},
  author={Tatarski, Valerian Ilich},
  year={2016},
  publisher={Courier Dover Publications}
}

@article{wu2021reliable,
  title={Reliable model to estimate the profile of the refractive index structure parameter {($C_{n}^{2}$)} and integrated astroclimatic parameters in the atmosphere},
  author={Wu, Su and Wu, Xiaoqing and Su, Changdong and Yang, Qike and Xu, Jiangyue and Luo, Tao and Huang, Chan and Qing, Chun},
  journal={Optics Express},
  volume={29},
  number={8},
  pages={12454--12470},
  year={2021},
  publisher={Optical Society of America}
}

@article{beland1993propagation,
  title={Propagation through atmospheric optical turbulence},
  author={Beland, Robert R},
  journal={Atmospheric Propagation of Radiation},
  volume={2},
  pages={157--232},
  year={1993},
  publisher={The Infrared and Electro-Optical Systems Handbook}
}

@article{menter2003ten,
  title={Ten years of industrial experience with the {SST} turbulence model},
  author={Menter, Florian R and Kuntz, Martin and Langtry, Robin},
  journal={Turbulence, Heat and Mass Transfer},
  volume={4},
  number={1},
  pages={625--632},
  year={2003}
}

@article{wang2016using,
  title={Using an artificial neural network approach to estimate surface-layer optical turbulence at Mauna Loa, Hawaii},
  author={Wang, Yao and Basu, Sukanta},
  journal={Optics letters},
  volume={41},
  number={10},
  pages={2334--2337},
  year={2016},
  publisher={Optical Society of America}
}

@article{alqaraghuli2023road,
  title={The road to high data rates in space: {T}erahertz versus optical wireless communication},
  author={Alqaraghuli, Ali J and Siles, Jose V and Jornet, Josep M},
  journal={IEEE Aerospace and Electronic Systems Magazine},
  volume={38},
  number={6},
  pages={4--13},
  year={2023},
  publisher={IEEE}
}

@article{saha2022turbulence,
  title={Turbulence strength {$C_{n}^{2}$} estimation from video using physics-based deep learning},
  author={Saha, Ripon Kumar and Salcin, Esen and Kim, Jihoo and Smith, Joseph and Jayasuriya, Suren},
  journal={Optics Express},
  volume={30},
  number={22},
  pages={40854--40870},
  year={2022},
  publisher={Optica Publishing Group}
}

@article{han2022molecular,
  title={Molecular absorption effect: A double-edged sword of {t}erahertz communications},
  author={Han, Chong and Gao, Weijun and Yang, Nan and Jornet, Josep M},
  journal={IEEE Wireless Communications},
  volume={30},
  number={4},
  pages={140--146},
  year={2022},
  publisher={IEEE}
}

\vfill
  
\end{document}